\documentclass{article}
\usepackage[dvips]{graphicx}

\begin{document}

\title{Oscillations and temporal signalling in cells}

\author{G. Tiana$^1$, S. Krishna$^2$, S. Pigolotti$^3$, M. H. Jensen$^2$ and K. Sneppen$^2$\\
$^1$Department of Physics, University of Milano and INFN \\via Celoria 16, 20133 Milano, Italy.\\
$^2$The Niels Bohr Institute, Bledgamsvej 16, 2100 Copenhagen, Denmark.\\
$^3$IMEDEA, C/ Miquel Marqu\`es, 21, 07190 Esporles, Mallorca, Spain.}

\maketitle
\begin{abstract}
The development of new techniques to quantitatively measure gene expression 
in cells has shed light on a number of systems that display 
oscillations in protein concentration. 
Here we review the different mechanisms which can produce
oscillations in gene expression or protein concentration, using a
framework of simple mathematical models.
We focus on three eukaryotic
genetic regulatory networks which show ``ultradian" oscillations, with time
period of the order of hours, and involve, respectively,
proteins important for development (Hes1), apoptosis (p53)
and immune response (NF-$\kappa$B). 
We argue that underlying all three is a common design consisting
of a negative feedback loop with time delay which is responsible
for the oscillatory behaviour.
\end{abstract}

\section{Introduction}

Biological systems display fascinating  spatial and temporal
patterns, which hint that the underlying
cellular processes 
are highly dynamic and operate on a wide range
of time and length-scales. 
This is indeed confirmed by measurements of the temporal dynamics
of protein concentrations and gene expression
levels for various signalling and response systems,
made using pulse-labelling \cite{Little1983}, $\beta$-galactosidase measurements 
and immunoblotting \cite{Sassanfar},
electromobility shift assays \cite{baltimore}, real time PCR \cite{RTPCR},
fluorescence techniques \cite{alon,GevaZatorsky,Nelsonetal,Bosisioetal}, chromatin immunoprecipitation
assays \cite{meticell}
and microarrays \cite{kuchocircadian}.
Overall, it is now evident that regulatory and signal transduction
networks do not depend merely on shifting the relevant protein concentrations
from one steady state level to another. Rather, the signals often have
a significant temporal variation that carries 
much more information and propogates through the regulatory networks
in a complex manner.
With time-resolved data now available for a number of response and
signalling systems, it is perhaps the appropriate 
time to explore whether there are any commonalities, or ``design principles", 
in the underlying mechanisms. In this review we will show that
a prominent subclass of such systems does indeed have a common underlying
design structure which combines a negative feedback loop with a time delay.

This subclass consists of systems that display
oscillatory behaviour under some conditions. 
The most obvious examples are circadian rhythms and
cell division.
Oscillations are also seen in the levels of cellular calcium \cite{calcoscreview}.
and in embryo development. Somite segmentation, for instance,
exhibits clearly periodic spatial patterns which are produced by
periodic temporal variation of proteins like Hes1, Axin,
Notch and Wnt \cite{hirata,pourquie,aulehla,dequeant}.
We also include in this class, systems which display
damped oscillations or semi-periodic behaviour. 
One example is oscillations, triggered by DNA damage,
in p53, a key protein involved in cell death and
apoptosis pathways \cite{oren,alon,Lahav,GevaZatorsky}. Hormones, such
as the Human Growth Factor, also show such
intermittently periodic behaviour and pulsatile secretion \cite{pulsehormone}. 
For these systems the recurrent behaviour
probably has a direct physiological role. In cyanobacteria, for example, various
physiological processes, like respiration and carbohydrate synthesis,
are directly influenced by the circadian clock to be in sync with 
the day-night cycle \cite{GIJK}.
For other systems, however, clear oscillatory behaviour is not observed in
the wild-type, but only in certain mutants. For instance, the NF-$\kappa$B
signalling system, involved in immune response in mammalian cells,
shows oscillations in nuclear NF-$\kappa$B concentration
only in a mutant which contains just one isoform
of the NF-$\kappa$B inhibitor, I$\kappa$B$\alpha$, and not the other isoforms
\cite{baltimore}.
Fluorescence measurements of the NF-$\kappa$B system modified so that I$\kappa$B$\alpha$
was overexpressed also showed sustained oscillations over several hours \cite{Nelsonetal}.
However, wild-type cells show at best damped oscillations \cite{baltimore}.
Here it is not clear if the oscillations themselves have a significant
physiological consequence or are merely a by-product of other
requirements for the wild-type behaviour.
Nevertheless, the sub-parts of these systems which are potentially oscillatory
are important, often essential, components which influence the
complex temporally varying wild-type response.

\begin{figure} 
\vspace{0.3cm}
\centerline{\includegraphics[width=8cm]{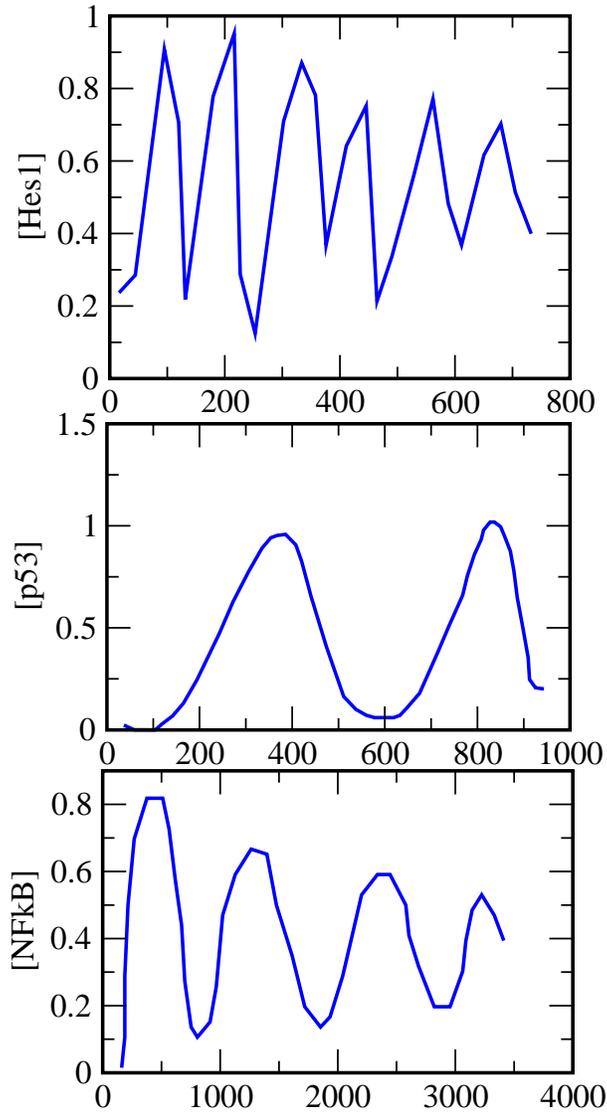}} 
\caption{Experimentally observed ``ultradian" oscillations in
(a) Hes1, data taken from ref.
\protect\cite{hirata}, (b) p53, data taken from ref.
\protect\cite{alon} and (c) NF-$\kappa$B, data taken from ref.
\protect\cite{baltimore}.} \label{fig_oscill} 
\end{figure}

We will focus on three of the regulatory systems mentioned above:
Hes1 in mammalian embryos, p53-Mdm2 in mammalian cells, and the
NF-$\kappa$B signalling system, also in mammalian cells.
Fig. \ref{fig_oscill} shows the oscillations observed in experiments for all three.
These systems are quite complicated, with many components interacting in various
ways -- including transcriptional activation/repression, 
translation and post-translation
regulation, protein-protein interaction, targeted protein degradation, and
active nuclear-cytoplasmic translocation -- composed into a complex network
with multiple feedback loops.
In this review, we mainly describe the mechanisms
and structures in these networks that
allow them to produce the observed oscillations.
We will do this using simplified mathematical models
where the level of description balances the need to correctly 
describe the systems with the need to coarse-grain over some details
in order to reveal common design features. 

We first begin by defining the overall framework of models we consider,
and describing the basic ingredients for producing
oscillatory behaviour within this framework --
{\em negative feedback} and {\em time delays}. 
In the subsequent sections we show how the three examples of Hes, p53 and
NF-$\kappa$B contain these ingredients. Therein we also elaborate, using stability
analyses, the requirements for producing oscillations in these systems.
Finally, we briefly
discuss how to extract information about underlying
structures from oscillatory time series data.

\section*{Negative feedback}

\begin{figure} \centerline{\includegraphics[width=12cm]{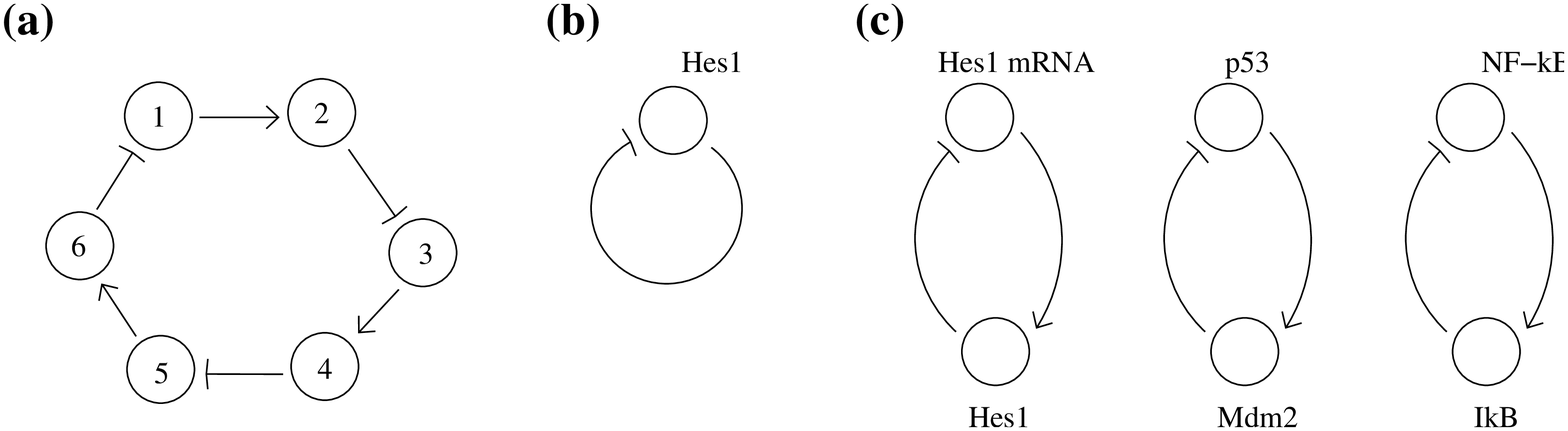}} \caption{Negative feedback loops. (a) A generic multi-species negative feedback loop, (b) the simplest negative feedback loop -- a self-repressor, (c) two component negative feedback loops for the three examples we examine.
In all figures a normal arrow represents an activating interaction, and a 
barred arrow represents a repressing interaction.} \label{schematic} \end{figure}

Fig. \ref{schematic}a shows a schematic negative feedback loop. The individual nodes in this
loop are the relevant dynamical variables: they could be protein concentrations,
gene expression levels, etc. Each variable either activates or represses the
next one in the loop. By an activator, we mean that if the concentration
of protein 1 goes up it tends to increase the concentration of protein 2
(either by increasing its production rate or decreasing its degradation rate).
A repressor has the opposite effect.
Then, a negative feedback loop is simply
defined as {\em a loop with an odd number of repressors}, so that
the effect of a perturbation in the concentration of any species 
in the loop eventually feeds back to
itself with a negative sign.
Feedback loops are, in general, the most common network motifs in cellular
organization, especially when one considers the regulation of
small molecules \cite{negloops}.
We concentrate on negative feedback here because it was hypothesized by 
Thomas \cite{thomas}, and rigorously
proven for a wide class of systems \cite{snoussi,gouze},
that the presence of at least one negative feedback loop is a necessary condition 
for oscillations.

The simplest negative feedback loop is of course a protein which represses
itself. There are many examples of such proteins: 
the main repressor of the SOS regulon in {\em E. coli}, LexA, also represses
its own production \cite{Schnarr}; Hes1, mentioned above, also represses transcription
of its own gene \cite{hirata}, shown schematically in Fig. \ref{schematic}b. 
However, Hes1 looks like a one-component loop only if we coarse-grain
over the intermediate steps in protein production. If we count the Hes1 mRNA
also then it becomes the two component negative feedback loop of Fig. \ref{schematic}c:
one component, Hes protein, represses the production of the second, 
Hes1 mRNA, while the
second component activates the first. 
The other two systems we discuss later in this review have the same loop structure
when coarse-grained to the two component level.
Nuclear NF-$\kappa$B is known to activate production of
I$\kappa$B$\alpha$, which inhibits nuclear import of NF-$\kappa$B by sequestering it
in the cytoplasm \cite{baltimore}. p53 and Mdm2 work similarly, with p53 activating the
{\em mdm2} gene and Mdm2 sequestering p53 \cite{Wuetal}.

To examine the possible dynamical behaviour produced by a negative
feedback loop, we model the dynamics of the concentrations of the
components using ordinary coupled differential equations. For instance,
for the two component loops shown in Fig. \ref{schematic}c:
\begin{eqnarray}
\frac{dx_1}{dt}&=&g_1(x_1,x_2)\nonumber\\ 
\frac{dx_2}{dt}&=&g_2(x_1,x_2),
\label{general} 
\end{eqnarray} 
where $x_1$ and $x_2$ are the concentrations of the two 
components $X_1$ and $X_2$.
This can easily be generalized to longer loops with $N$ components:
\begin{equation}
\frac{dx_i}{dt}=g_i(x_i,x_{i-1}),\quad i=1,2,\ldots,N,
\label{generalN}
\end{equation}
which models a single feedback loop with no cross-links because
the rate of change of a given variable $x_i$ depends
only on itself and the preceding variable, $x_{i-1}$.
In writing such an equation we are assuming that fluctuations
in space and time are negligible. Thus, there are no stochastic or
diffusion terms.
The functions $g_i$ model both production and degradation of the 
components, and can take many forms depending on the 
kind of interactions in the system.
For example, in the p53 example (with $X_1 \to$ p53, $X_2 \to$ Mdm2), we know that p53 binds
to an operator site at the promoter for the {\em mdm2} gene and aids
transcription. Then, under some assumptions 
\footnote{In particular, this assumes that the number of proteins bound
to the operator site is much smaller than the total number of proteins.}
the production of Mdm2 can be modeled using a term of
the form:
\begin{equation}
\frac{(p/K)^h}{1+(p/K)^h},
\label{MM}
\end{equation}
a sigmoidal monotonically
increasing function of
the p53 concentration, $p$ (the Hill coefficient, $h\ge 1$, accounts for cooperativity in
the binding of p53 to the operator).

It is difficult to say anything about the behaviour of the general 
coupled differential equation of a negative feedback loop, like Eq. \ref{generalN}, 
However, it is reasonable to constrain the functions $g_i$ of 
Eq. \ref{generalN} to be monotonic in $x_{i-1}$. This corresponds to saying that
a protein that activates a particular process cannot change to repress it at some
other concentration, and vice versa, which is the case for most transcription factors
\footnote{Some proteins can both activate
as well as repress the same process depending on their concentration.
For instance, CI in
lambda phage activates the $P_{\mathrm{RM}}$ promoter at low concentrations, but represses it at high
concentrations
\cite{CIregulation2}.  Another example is the galactose
regulator GalR, which at high concentrations of galactose activates 
the promoter {\sl galP2} but in the absence of galactose
forms a DNA loop, which completely represses {\sl galP2}
\cite{semseyvirnik}. Such examples are, however, somewhat rare and we
will not consider them.
}. 
For monotone systems, not only is there no ambiguity about whether the loop implements
negative or positive feedback, but we can also prove rigorously that there is only
one fixed point (see Appendix A).
The question then is whether such a steady state is stable or unstable.
For two variables, if the fixed point is unstable, and the system's 
trajectories are bounded (again, a reasonable assumption for 
a biological system) then it
must show periodic oscillations: this is known as the Poincar\'e-Bendixson theorem \cite{strogatz}.
For monotone systems this is true even when the loop has more than two
components \cite{nochaosone}, which means that the question of whether oscillations are
possible boils down to whether the fixed point is stable or unstable.

\section*{Time delays}

Physically, what is required for instability of the fixed point, and hence
oscillations, is a time delay, or a
slowing down of the signal going round the loop. 
By signal we mean perturbations of the concentrations
away from the steady state. If a perturbation
in the concentration of one variable instantaneously affects the concentration
of the next one, and so on, then for a negative feedback loop, any perturbation
will be immediately cancelled and the steady state will be stable. 
A sufficiently large time delay on the other hand will produce oscillations.
This can be readily understood with an example: 
consider a person who walks along a
straight line to a given point, marked on the ground. If this person is able to
take instantaneous decisions, he will approach the mark and then stop. This is
a stationary solution to the walk kinetics. If, on the other hand,  it takes
some time to realise that the mark has been reached, the person will not stop
at the mark, but cross it. When eventually the information that the mark has
been crossed is processed, the person will turn back and walk in the opposite
direction. The mark will again be reached and overshot, and so on. The
resulting kinetics will be damped or sustained oscillations about the mark.
In cellular systems many processes could produce time delays. Table 1 lists
some timescales associated with such processes.

With the above insight, one can see that
a simple way to model oscillations using the framework of Eq. \ref{general} 
is to introduce an explicit time delay into the equations:
\begin{eqnarray}
\frac{dx_1(t)}{dt}&=&g_1(x_1(t),x_2(t),x_1(t-\tau_1),x_2(t-\tau_2))\nonumber\\
\frac{dx_2(t)}{dt}&=&g_2(x_1(t),x_2(t),x_1(t-\tau_3),x_2(t-\tau_4)),
\label{general_delay} 
\end{eqnarray} 
Of course, this is the most general form, and often it is
possible to have fewer than four delays.
Oscillations are observed even in
the simplest case of a linear differential equation with
a single (sufficiently large) delay (see Appendix C): 
$dx/dt=-x(t-\tau) \label{eq_linear}$, which models the 1-component Hes loop
of Fig. \ref{schematic}b where Hes1 represses itself after a time delay $\tau$.
Although a linear delay rate equation is an oversimplification of Hes1
production, the physics which lies behind more general delay differential 
equations like Eq. (\ref{general_delay}) is the same. The added complication
is that the functions $g_1$ and $g_2$ are in general highly 
nonlinear, resulting in an amplification of the effect of the delay.

Putting an explicit time delay like this, of course, 
does not really shed light on the mechanism producing
the time delay. There are several possibilities which can
be used to produce oscillations in a negative feedback loop:\\
\begin{enumerate}
\item a process that takes a finite minimum time
\item many intermediate steps
\item a sharp response by some of the variables
\item saturated degradation
\item autocatalysis\\
\end{enumerate}

\noindent
To elaborate:\\

\begin{enumerate}
\item Rate equations, like Eq. \ref{generalN}, typically model
processes which occur with a given average rate, such as the binding
of a protein to an operator site. A hidden assumption
is that the time interval between two binding events is
Poisson distributed, which means that often there is a reasonable
probability for two events to be separated by a very short 
time interval (say, much
shorter than the average time between events).
Sometimes, however, molecular processes
take a certain minimum time. For instance, if transcription and translation
take a time $\tau$ after a polymerase binds to the promoter,
then the rate of production of the protein is more appropriately modeled
as $dx/dt\propto P(t-\tau)$, where $P(t)$ is 1 if the polymerase is bound
to the promoter at time $t$, and zero otherwise. Such logic has been
used to justify time-delay models in a variety of systems \cite{nhes1,tiana,lewis}. 
This is the approach we will take to model the p53 and Hes
systems, discussed in the subsequent sections.\\

\item Processes like transcription and translation have this character because
they are in fact composed of a large number of intermediate processes:
the polymerase binds, first forming a closed complex, which then makes a
transition to an open complex, and then to an elongating complex, followed
by many ``steps" along the DNA until the polymerase reaches the end of the
gene. Even if each of these individual steps is a Poisson process, the net
effect adds up to a time delay. Thus, instead of putting in an explicit
time delay as in Eq. \ref{general_delay}, one could work with a negative
feedback loop with many components.
One simple example of such an oscillator is the repressilator \cite{repressilator},
which is a negative feedback loop with six components.\\

\item The repressilator also has
another necessary ingredient -- a nonlinearity in the $g_i$ functions which allows
some of the variables to respond to changes in the preceding variable 
in a faster-than-linear fashion. More precisely, the repressilator uses sigmoidal
functions, like the function \ref{MM}, to describe transcriptional repression, 
and needs Hill coefficients of at least 2 in order to achieve oscillations. The earliest 
example of a negative feedback oscillator which uses a nonlinearity like
this to produce a sharp response is a model by Goodwin \cite{Goodwin}
where the Hill coefficient needs to be more than 8. 

\item Such high Hill coefficients are unlikely for biological systems, however.
In order to get around this ``problem", Bliss, Painter and Marr \cite{BPM} 
introduced
another way of producing an effective time delay. They used the saturated
degradation of one of the concentrations. Saturated degradation means that 
there is an upper limit to the degradation rate of one species,
thereby allowing it to remain abundant for a longer time, thus effectively
slowing down the signal travelling around the loop. Such saturated degradation
is quite common in biological systems, especially when proteins are
tagged for targeted degradation by another protein, as we will
show in the NF-$\kappa$B case discussed below.

\item Finally, autocatalysis, where a molecule activates its own production
can be used to produce oscillations in systems like Eq. \ref{general} \cite{ploscomp}. 
In fact for two variable systems where there is no explicit time-delay
this is a necessary condition for oscillations \cite{conradthesis}. Note that
this modification typically makes the system non-monotonic.
\end{enumerate}
  
This survey of the theoretical requirements for producing oscillations
in negative feedback loops already allows us to make an interesting
observation: Monotone 2-component loops without an
explicit time-delay cannot oscillate, whatever the nonlinearity in
the $g_i$ functions. We prove this explicitly in Appendix A.
Thus, if one insists on modelling an oscillating
system using two variables, one must choose between introducing a time
delay, and sacrificing monotonicity.

\begin{table} 
\begin{tabular}{|l|l|}\hline 
translocation through nuclear pores \protect\cite{pores} & $10^{-4}$s \\ 
diffusion in eucaryotic cell & $1$ s\\ 
translation \protect\cite{alberts} & 30 s \\ 
transcription \protect\cite{alberts} & 3 min \\ 
mRNA degradation \protect\cite{alberts} & 3 min \\\hline 
protein degradation & 10 min to 10 h \\ cellular signals
\protect\cite{oren,hirata,baltimore} & 1h\\ \hline 
\end{tabular} 
\caption{Time scales of some cellular processes associated with a single
molecule. The upper part of the table indicate the processes which are usually
neglected in writing the rate equations of regulation newtworks, while the
lower part indicate proceeses which are usually accounted for.}
\label{tab_timescales} \end{table}

\section{Ultradian oscillations in biological systems}

\begin{figure} \centerline{\includegraphics[width=10cm]{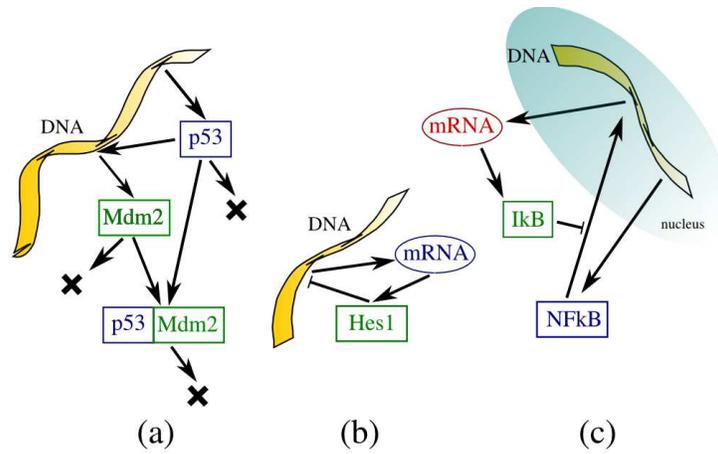}} \caption{A
sketch of the feedback loops controlling the concentration of p53
(a), Hes1 (b) and NF-$\kappa$B (c).} \label{fig_sketch} \end{figure}

We now turn to three biological
systems to illustrate these ideas in action. 
In the following, we will briefly
give a description of the systems, p53-Mdm2, Hes1 and NF-$\kappa$B, in which ``ultradian"
oscillations have been observed, which have time periods of the
order of hours, as opposed to ``circadian" 24 hour rhythms.
We will discuss specific details of each system, at the same time 
emphasising that the basic physical mechanism which produces oscillations in all three is
the same -- negative feedback along with time delays. 

\subsection{p53-Mdm2}

The protein p53 is responsible for inducing apoptosis in cells with damaged
DNA \cite{lane93}. The concentration of p53 is usually kept low by a feedback
mechanism involving another protein, Mdm2, which binds to p53 and promotes its
degradation. When the DNA is damaged, the cell expresses a number of kinases
which phosphorylate SER20 in p53, changing its affinity to Mdm2. This results
in oscillations in the concentration of p53, observed both in western blot
analysis \cite{oren,baror} (cf. Fig. \ref{fig_oscill}(a)) and in 
single cell fluorescence experiments \cite{alon,GevaZatorsky}. The standard
explanation for the overall increase in the concentration of p53 is
that its phoshoprylation decreases its affinity to Mdm2, shifting the
thermodynamic equilibrium towards higher concentrations.

Apart from not explaining the oscillations, this argument does not agree
with other experimental evidence. Equilibrium isothermal titration
calorimetry experiments have shown \cite{fersht} that 
phosphorylation at SER20 {\em decreases}, not increases, the dissociation
constant between p53 and Mdm2 from $k_D=575\pm 19\; nM$ to $k_D=360\pm 3\; nM$.
The same effect is observed {\it in vivo} \cite{gottlieb}, where p53ASP20 (a
mutated form which mimics phosphorylated p53) binds Mdm2 more tightly than
p53ALA20 (which mimics unphosphorylated p53). Moreover, single cell
experiments \cite{alon} show a slight decrease in the concentration of p53
after DNA damage, which cannot be explained by the standard argument.

\begin{figure} 
\vspace{0.3cm}
\centerline{\includegraphics[width=10cm]{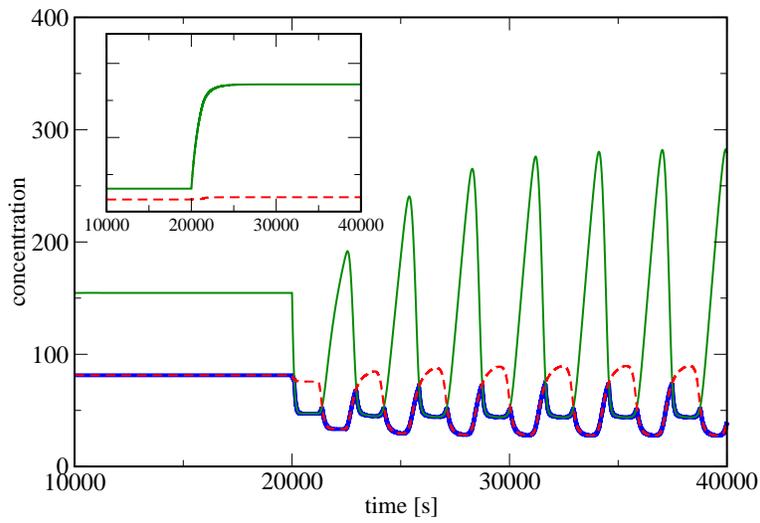}}
\caption{Oscillations displayed by the numerical solution of the dynamic
equations of p53. The numerical values of the rates are $a=3\cdot
10^{-2}\;s^{-1}$, $b=10^{-4}\;s^{-1}$, $c=1\;s^{-1}$, $d=1\;s^{-2}$,
$S=1\;s^{-1}$, $k=180$, $k_g=28$ and $\tau=1200\;s$. At time $t=20000$ s the
value of $k$ is decreased of a factor 10. In the inset, the solution of the
same equations, where the value of $k$ is increased of a factor 10 at $t=20000$
s. The equations are solved with the Adams algorithm. } \label{fig_p53o}
\end{figure}

We showed that a simple model of the p53-Mdm2 system that incorporates the time delay
associated with some relatively slow processes within the cell can account 
for the experimental facts in a simple way \cite{tiana}.
The feedback mechanism is
sketched in Fig. \ref{fig_sketch}(a) and the associated time-delayed rate
equations are \begin{eqnarray} \label{eq_p53} \frac{\partial p}{\partial t} & =
& S - a\cdot pm - b\cdot p \\\nonumber \frac{\partial m}{\partial t} & = &
c\frac{p(t-\tau)-pm(t-\tau)}{k_g+p(t-\tau)-pm(t-\tau)} - d\cdot m \\\nonumber
pm & = & \frac{1}{2}\left((p+m+k)-\sqrt{(p+m+k)^2-4p\cdot m}\right),
\end{eqnarray} where the delay $\tau$ takes into account the half-life of
mRNA, the diffusion time, the time needed to cross the nuclear membrane and the
transcription/translation time. One can solve Eqs. (\ref{eq_p53}) numerically,
simulating the damage to DNA by a sudden change in the dissociation constant
$k$. Fig. \ref{fig_p53o} shows the onset of oscillations in this model in
response to a sudden decrease in $k$ at time $t=2000\;s$. 

Several observations emerge from the simulation results in Fig. \ref{fig_p53o}.
Most interestingly, what triggers oscillations is a decrease in the dissociation constant $k$,
while any increase in its value just shifts the equilibrium towards higher
values of $p$ (cf. inset of Fig.
\ref{fig_p53o}). Moreover, just after $t=2000\;s$ the concentration of p53
decreases, as observed in the experiments. Finally, the concentration, $pm$, of
the complex p53-Mdm2 is, at any time, essentially identical to the minimum
of $p$ and $m$ (gray curve in Fig. \ref{fig_p53o}), indicating that the
complex is saturated.

\begin{figure} \centerline{\includegraphics[width=8cm,angle=-90]{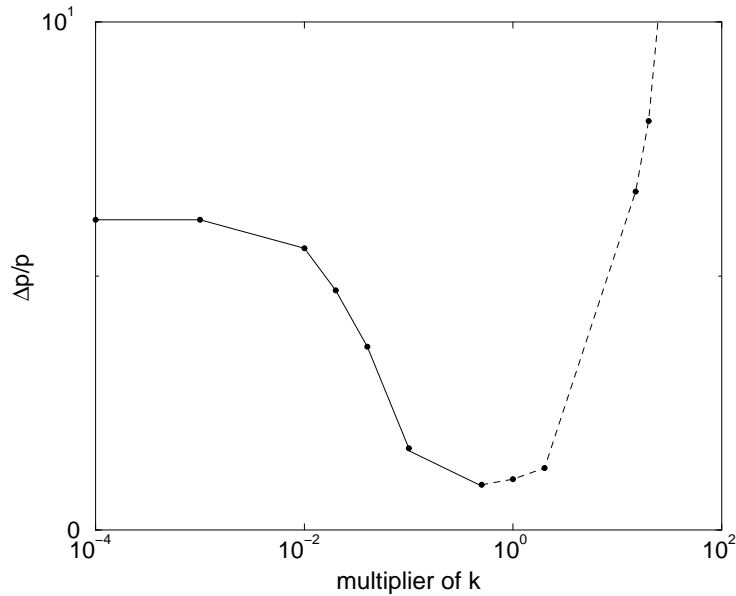}}
\caption{The height of the response peak $\Delta p$ with respect to the
quantity that multiplies $k$, mimicking the stress. The dotted line indicates
that the system does not display oscillatory behaviour.} \label{fig_p53k}
\end{figure}

The relative change, $\Delta p$, in the maximum concentration of p53 reached
after the simulated stress event is displayed in Fig. \ref{fig_p53k} as a
function of the change in $k$. The value of $\Delta p$ displays a sigmoidal 
behaviour if the stress decreases the value of $k$.
In contrast, in the absence of a time delay, this curve is linear with
respect to $k$ \cite{tiana}. Thus, the time delay and the oscillations are
crucial for producing a sensitive system that can respond to stress in a
sharp, faster-than-linear manner.

We also undertook a detailed analysis of the response of the oscillations to changes in the
parameters \cite{tiana}. It turns
out that the oscillations do not change qualitatively when 
parameters $a$, $b$ and $c$ are varied (by upto five orders of magnitude), whereas
a decrease in $d$ or $k_g$ suppresses the oscillating behaviour. 
These parameters are associated with, respectively, the 
the degradation of Mdm2 and the binding of p53 to
the DNA. One could speculate therefore that these processes play an important
role in the production of tumors that arise when the p53 response mechanism fails.
This speculation is, in fact, consistent with the observation that
around the 45\% of all tumors display mutations
in the p53 region which binds to DNA \cite{greenblat94}.

\subsection{Stability analysis of the p53-Mdm2 model}

The overall behaviour of the proteins with respect to time evidently depends on
the parameters of the dynamic equations, that is the production and degradation
rates and the delay. Over long timescales, the protein concentrations can
either converge to a steady state or a limit cycle (i.e., sustained oscillations).
Chaotic behaviour is never observed within this model.

A careful analysis of how the dynamics of the protein concentration depends on
the parameters of the rate equations has been done by Neamtu and coworkers in
ref. \cite{neamtu}. The main conclusion of this work is that under the
condition that the dissociation constant $k$ between p53 and Mdm2 is small,
there exists a critical delay, $\tau_0$, above which the system shows 
oscillations (see details in Appendix C).

In the language of dynamical
systems, the transition when
the delay, $\tau$, crosses $\tau_0$ is a Hopf bifurcation.
A Hopf bifurcation is the generic type of bifurcation that occurs when a
stable fixed point of the system becomes unstable and turns into a limit cycle,
a dynamically oscillating state, as consequence of a change in
some parameter of the system (see Appendix B for more details).  

\subsection{Hes1 and its mRNA}

The transcription factor Hes1 controls the differentiation of neurons in mammalian
embryos \cite{kageyama97}. Its concentration is controlled by a feedback loop
built out of Hes1 and its own mRNA (see Fig. \ref{fig_sketch}). Like p53, its
concentrations displays oscillations when the cells are stimulated with serum
\cite{hirata}. The period is similar to that of p53, 
approximately two hours, and the oscillations last for $\approx 12$ hours. 
Hes1 is known to repress another protein called Mash1.
Both the knocking out of Mash1 and its continuous expression by means of
retroviral introduction results in a lack of cell differentiation. 
Only when there are periodic oscillations in the concentration of Hes1 (and thus of Mash1) 
is there proper differentiation of neuronal cells \cite{kageyama97}, showing that
the temporal variation of the protein concentrations are 
critical for the developement of the nervous system.  

The relation of Hes1 oscillations to segmentation and spatial
patterns has been studied by several authors. It is well known
that oscillations in Hes1 is part of Notch signalling and
Hirata et al Ref. \cite{hirata} studied the coordinated somite segmentation in
the presomitic mesoderm of mice embryos and found a correlation
between the oscillations of Hes1 and initiation of somites.
The general issue of segmentation in vertebrates has been further
studied in Ref. \cite{pourquie,aulehla,dequeant} indicating that the
oscillations in the Notch pathway signaling are intimately
related to the Wnt signaling. At the caudal end of the embryo
Notch and Wnt oscillates out of phase when the gradient in the Wnt
level is over a given threshold \cite{aulehla,dequeant}. New cells are
provided and move away from the caudal
end each setting a boundary for a somite segmentation. This
continues until all somites are produced which subsequently gives
rise to the spinal cord \cite{pourquie}.
Ultradian oscillations of signaling
pathways are thus extremely important for segmentation.

The control mechanism of Hes1 oscillations again involves a negative feedback loop with
one activation and one repression: the
transcription of the mRNA of Hes1 activates the production (translation) of the protein,
and Hes1 represses the transcription
of its own mRNA. The main difference compared to p53 is that here one of the
nodes represents an mRNA, not a protein.
However, this difference is merely semantic:
the control network still has two nodes, of which one is
activating and one is repressing. The molecular species of these nodes are
immaterial to the description of the oscillations
\footnote{We could of course introduce two more nodes in the p53 network
representing the mRNA of p53 and Mdm2. However, this would not change the
logic of the loop, replacing an activation by two successive activations.
As discussed earlier adding more intermediate nodes introduces a time delay.
Since the p53 equations had an explicit delay it seems redundant to add the
mRNA nodes also.}.

We model the system using the following equations (cf. \cite{nhes1})
\begin{eqnarray} \frac{\partial r}{\partial t} & = &
\frac{\alpha k^h}{k^h +[s(t-\tau)]^h} -k_r r(t)\\ \frac{\partial s}{\partial t}
& = & \beta r(t) - k_s s(t), \label{eq_hes1} \end{eqnarray} where
$s$ and $r$ are the concentrations of Hes1 and its mRNA, respectively. The
meaning of these equations is that mRNA is produced at a rate $\alpha$ when Hes1
is bound to the DNA. The probability that Hes1 is bound to DNA is
$k^h/(k^h+s^h)$, where $k$ is a characteristic concentration for dissociation
of Hes1 from the DNA, and $h$ is the Hill coefficient that takes into account
the cooperative character of the binding process.
$k_{r}$ and $k_{s}$ are the spontaneous degradation rates of the two
proteins, while $\tau$ is the delay associated with the molecular processes
that we do not want to describe explicitly (transcription, translocation,
etc.).  Ref. \cite{hirata} suggests that $\tau_{rna}$ and $\tau_{hes1}$
are of the order of 25 minutes.  The value of the time delay is difficult to
assess, since it is determined by a combination of various molecular processes. One can
guess that its order of magnitude is tens of minutes. 

\begin{figure} 
\vspace{0.3cm}
\centerline{\includegraphics[width=10cm]{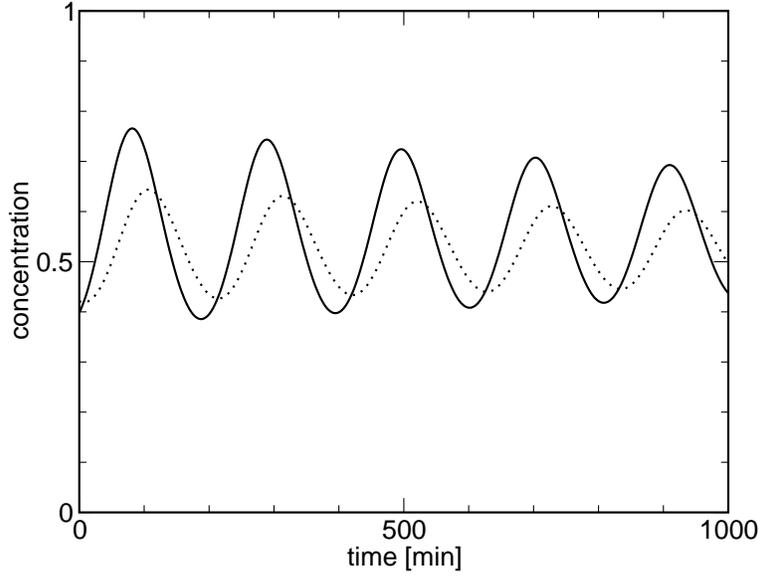}} 
\caption{The
concentration of Hes1 (solid curve) and uts mRNS (dashed curve) obtained
solving numerically Eq. (\protect\ref{eq_hes1}). } \label{fig_hes1}
\end{figure}

\begin{figure} \centerline{\includegraphics[width=10cm]{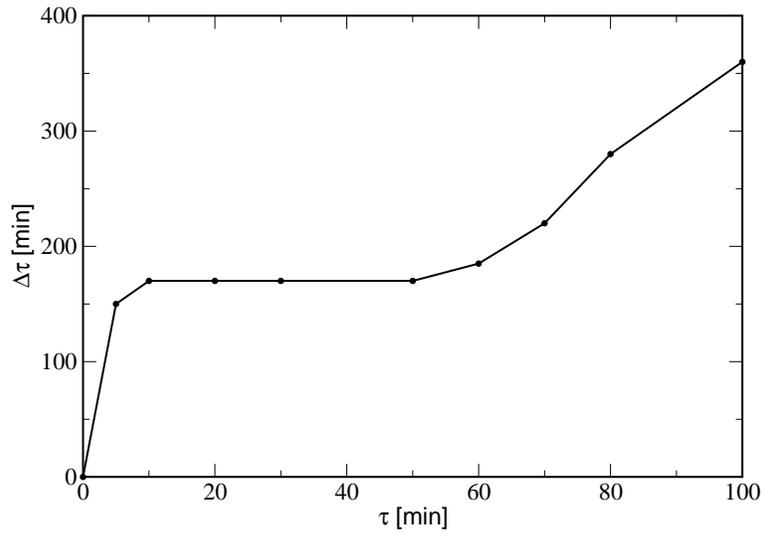}} \caption{The
oscillation period $\Delta\tau$ of Hes1 as a function of the delay $\tau$. }
\label{fig_hes1del} \end{figure}

The numerical solution of
Eq. (\ref{eq_hes1}) is displayed in Fig. \ref{fig_hes1}. The oscillations have
a period $\Delta\tau\approx 170$ min. The dependence of the time period on the delay
$\tau$ is shown in Fig. \ref{fig_hes1del}. For any delay in the range
$10<\tau<50$ min, the oscillation period is consistent with that found
experimentally, and so is the time difference between the peaks in Hes1 and mRNA, $\approx 18$ min.
For $\tau<10$ min, the
system shows no oscillations. To check the robustness of the results, we have
varied $\alpha$, $\beta$ and $k$ over 5 orders of magnitude around the basal
values, listed in the caption to Fig. \ref{fig_hes1}, and observed no
qualitative changes to the oscillatory behaviour. On the
other hand, a decrease of $k_{s}$ and $k_{r}$ disrupts the oscillations.
This is because these two quantities set the timescale of the dynamics,
Increasing this timescale, keeping
$\tau$ constant, is equivalent to decreasing $\tau$ and
thus no oscillations are produced.

The behaviour of Hes1 is very similar to that of p53, both in the features
of the oscillations and in the lag delay before they start. This is
not unexpected, since the structure of Eq. (\ref{eq_hes1}) is very similar to
the structure of Eq. (\ref{eq_p53}) and to any time delay equation describing a two component
negative feedback loop.

\subsection{NF-$\kappa$B and I$\kappa$B} \label{sect_nfkb}

The NF-$\kappa$B family of proteins is one of the most studied in the last ten
years, being involved in a variety of cellular processes including immune
response, inflammation, and development \cite{baltimore,hoffmanncell}.  
NF-$\kappa$B can be activated by a
number of external stimuli \cite{Pahl} including bacteria, viruses and various
stresses and proteins (for instance, refs. \cite{baltimore,Nelsonetal} used the 
tumor necrosis factor-$\alpha$, TNF-$\alpha$).
In response to these signals it targets over 150 genes including many chemokines,
immunoreceptors, stress reponse genes, as well as acute phase inflammation
response proteins \cite{Pahl}.
Each NF-$\kappa$B has a partner inhibitor called I$\kappa$B, which inactivates
NF-$\kappa$B by sequestering it both in the nucleus as well as in the
cytoplasm. In fact, the I$\kappa$B proteins come in several isoforms
$\alpha,\beta,\epsilon$ \cite{baltimore,hoffmannmodel}, 
and perhaps others \cite{hoffmanncell}. 
Some of these isoforms are, in turn, transcriptionally
activated by NF-$\kappa$B, thus forming a negative feedback loop which is
essentially identical in structure to the other two discussed above.

The potential for this negative feedback loop to produce oscillations in the
nuclear-cytoplasmic translocation of NF-$\kappa$B was initially shown by
electrophoretic mobility shift assay experiments \cite{baltimore}.
They found that wild-type cells and mutants
containing only the I$\kappa$B$\alpha$ isoform showed damped oscillations.
In contrast, cells with only the I$\kappa$B$\beta$ or $\epsilon$ isoforms do
not show oscillations.  This conclusion was bolstered by single-cell
flourescence imaging experiments which show sustained oscillations of nuclear NF-$\kappa$B
in mammalian cells \cite{Nelsonetal}, with a time period
of the order of hours. In these experiments I$\kappa$B$\alpha$ was
overexpressed hence the system behaves like the mutant which has only the
I$\kappa$B$\alpha$ isoform.

\begin{figure} 
\centerline{\includegraphics[width=10cm]{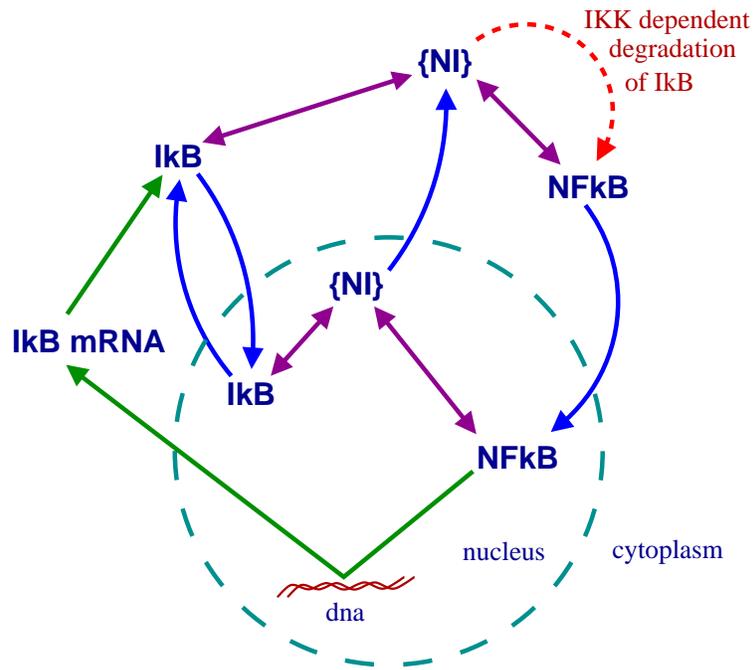}} 
\caption{The important interactions in the NF-$\kappa$B system. Green arrows indicate
transcription and translation. Blue arrows indicate transport processes. Purple
double arrows indicate complex formation. The red dashed arrow indicates IKK triggered
degradation of I$\kappa$B when complexed to NF-$\kappa$B.
\label{nfkbschematic}} \end{figure}

As sustained oscillations are clearest here, we will focus our modelling
on this mutant. The following cellular processes,
summarized in Fig. \ref{nfkbschematic}, are important for this system on the
timescales we are interested in (i.e., we ignore processes which are very
slow):\\ 
\begin{itemize} \item NF-$\kappa$B, when in the nucleus, activates
transcription of the I$\kappa$B$\alpha$ gene (henceforth we will drop $\alpha$
unless we are explicitly talking about more than one isoform), producing
I$\kappa$B mRNA in the cytoplasm.  \item The I$\kappa$B mRNA is translated to
form I$\kappa$B protein.  \item The I$\kappa$B protein can be transported in
and out of the nucleus.  \item In both compartments, I$\kappa$B forms a complex
with NF-$\kappa$B.  \item The NF-$\kappa$B-I$\kappa$B complex (henceforth
referred to as \{NI\}) cannot be imported into the nucleus. However, if it
forms within the nucleus it can be exported out.  \item Free NF-$\kappa$B
behaves in exactly the opposite way. Free NF-$\kappa$B is actively transported
into the nucleus but not from the nucleus to the cytoplasm.  \item The
cytoplasmic \{NI\} complex is tagged by another protein, the I$\kappa$B kinase
(IKK), for proteolytic degradation. This results in degradation of I$\kappa$B
only, releasing NF-$\kappa$B. Note that this degradation does not occur for
free I$\kappa$B.  \end{itemize}

On the timescales of interest, there is no net production or
degradation of NF-$\kappa$B. It simply cycles in and out of the nucleus, i.e.,
the sum of nuclear and cytoplasmic NF-$\kappa$B concentrations is a constant.
Amongst the above listed processes, the association and dissociation of the
complex \{NI\} occurs fast enough that the concentration of the complex can be
taken to be always in equilibrium with the free NF-$\kappa$B and I$\kappa$B
concentrations. This allows us to describe the system using a very simple model
consisting of only three variables \cite{sandeeppnas}, nuclear NF-$\kappa$B
($N_n$), cytoplasmic I$\kappa$B ($I$) and I$\kappa$B mRNA ($I_m$):
\begin{eqnarray}
\frac{dN_n}{dt}&=&A\frac{(1-N_n)}{\epsilon+I}-B\frac{IN_n}{\delta+N_n},\label{nfkb1}\\
\frac{dI_m}{dt}&=&N_n^2-I_m,\label{nfkb2}\\
\frac{dI~~}{dt}&=&I_m-C\frac{(1-N_n)I}{\epsilon+I} \label{nfkb3}.
\end{eqnarray}

The terms in equation (\ref{nfkb1}) model the nuclear import and export of
NF-$\kappa$B. Eq. (\ref{nfkb2}) models NF-$\kappa$B-activated
transcription of the I$\kappa$B$\alpha$
gene and spontaneous degradation of the mRNA.
Finally, eq. (\ref{nfkb3}) has terms for translation of I$\kappa$B mRNA
into the protein and IKK mediated degradation of I$\kappa$B.
The external signal is supplied by IKK that enters the equations through the
parameter, $C$, which is proportional to IKK concentration. 

\begin{figure} \centerline{\includegraphics[width=10cm]{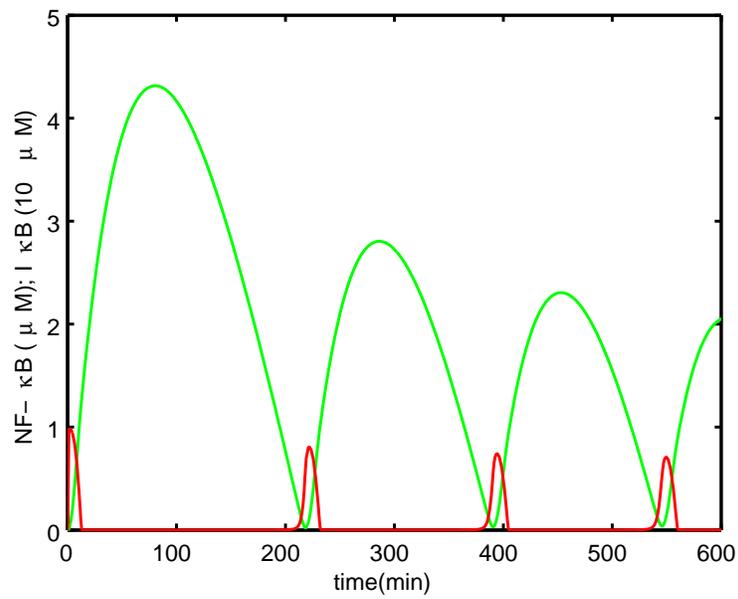}}
\caption{Oscillations of nuclear NF-$\kappa$B ($N_n$), red, and cytoplasmic
I$\kappa$B, green, for $A=0.007$, $B=954.5$, $C=0.035$, $\delta=0.029$ and
$\epsilon=2\times 10^{-5}$ (these parameter values are derived from the ones
used in ref. \protect\cite{baltimore}, see \protect\cite{sandeeppnas}.) In order to
facilitate comparison with the experimental plot, the x-axis has been limited
to 600 minutes, but the oscillations are in fact sustained (see Fig. \ref{spikyandsoft}a).} \label{osc}
\end{figure}

These equations produce sustained oscillations in all variables
over a wide range of parameter values. 
Fig. \ref{osc} shows the result of simulations which use 
parameter values from ref. \cite{baltimore}. The oscillations produced by
this model match the experimentally observed features well, in particular,
the shape, phase, time period and frequency response are correctly
reproduced \cite{sandeeppnas}.

Note that this is certainly not the only model able to reproduce the
experimental oscillations observed in NF-$\kappa$B.  Hoffman {\it et al.} have
constructed a long list of chemical reactions between 26 different molecules in
the NF-$\kappa$B system, including 65 numerical parameters 
\cite{baltimore,hoffmannmodel}.
The three variable model above was formed by reducing this larger model \cite{sandeeppnas},
also producing a 7-variable and 4-variable model in the process.
Hayot and Jayaprakash have also built a model with seven variables for NF-$\kappa$B oscillations \cite{HJ}.
Another model, which is similar to Hoffman et al's model, but has an additional feedback loop is
described in \cite{Lipniackietal}.

\subsection{Saturated degradation of I$\kappa$B}

A key element in the model  is the saturated degradation of cytoplasmic
I$\kappa$B  in the presence of IKK (second term in Eq. (\ref{nfkb3})).
This saturation occurs because the level of the NF-$\kappa$B-I$\kappa$B 
complex saturates, and this complex is needed
for IKK triggered degradation of I$\kappa$B. 
A stability analysis of the system shows the importance of the saturated
degradation for oscillations. We begin by examining the fixed points of the
system.

The fixed point values of $N_n$, $I_m$ and $I$ are solutions to
$$A\frac{(1-N_n)}{\epsilon+I}-B\frac{IN_n}{\delta+N_n}=0,$$ $$N_n^2-I_m=0,$$
$$I_m-C\frac{(1-N_n)I}{\epsilon+I}=0.$$
                                                                                                                                                                                                                                             
$I_m$ and $I$ can be eliminated using $I_m=N_n^2,$ giving
$I=(N_n^2\epsilon)/(C-CN_n-N_n^2).$
From this we find that the fixed point value of $N_n$ is a solution of the
equation $(C-CN_n-N_n^2)^2=(BC\epsilon^2N_n^3)/[A(\delta+N_n)],$
or equivalently,
$$N_n^5+(\delta+2C)N_n^4+C\left[2(\delta-1)+C-\frac{B}{A}\epsilon^2\right]N_n^3+C[(C-2)\delta-2C]N_n^2+C^2(1-2\delta)N_n+C^2\delta=0.$$
                                                                                                                                                                                                                                             
In general, this has two real solutions, one with $C-CN_n-N_n^2>0$ and the
other with $C-CN_n-N_n^2<0$. The latter results in a negative value for $I$ and
therefore is not an acceptable solution. Thus we are left with only one fixed
point.

Next, we linearize the equations around the fixed point, which gives the
Jacobian (see also Appendix B) 
$$J=\left( \begin{array}{ccc} -\frac{A}{\epsilon+I}-\frac{\delta
BI}{(\delta+N_n)^2} & 0 &
-\frac{A(1-N_n)}{(\epsilon+I)^2}-\frac{BN_n}{\delta+N_n} \\ 2N_n & -1 & 0\\
\frac{CI}{\epsilon+I} & 1 & -\frac{C\epsilon(1-N_n)}{(\epsilon+I)^2}\\
\end{array} \right)$$

\begin{figure} \centerline{\includegraphics[width=10cm]{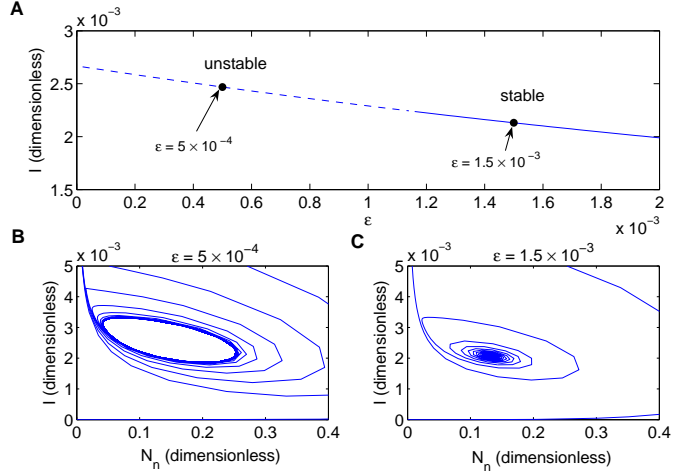}}
\caption{(A) The stationary value of $I$ as a function of the parameter
$\epsilon$ which controls the binding between NF-$\kappa$B and IKK in the
cytoplasm. (B) and (C) The projection of two trajectories at different values
of $\epsilon$ onto the $N_n-I$ plane.} \label{epsstability} \end{figure}

This matrix can be used to examine the stability of the fixed point:
it is unstable if any of the eigenvalues have a positive real part.
For the NF-$\kappa$B system, Fig. \ref{epsstability} shows how the
stability depends on the parameter $\epsilon$. When $\epsilon$ is
small compared to the steady state value of $I$, the degradation rate
of I$\kappa$B is independent of $I$, which is what we mean by the term
``saturated degradation". On the other hand, when $\epsilon$ is large
then the degradation is proportional to $I$ and is not saturated.
Fig. \ref{epsstability} shows that, indeed, the fixed point is unstable
when $\epsilon$ is small compared to $I$. 
Physically, this is because saturated degradation
introduces an effective time-delay into the feedback--loop:
it allows I$\kappa$B to accumulate and stay around longer than with
non-saturated degradation.

\subsection{Spikiness and sharp response of the NF-$\kappa$B oscillation}

\begin{figure} 
\centerline{\includegraphics[width=10cm]{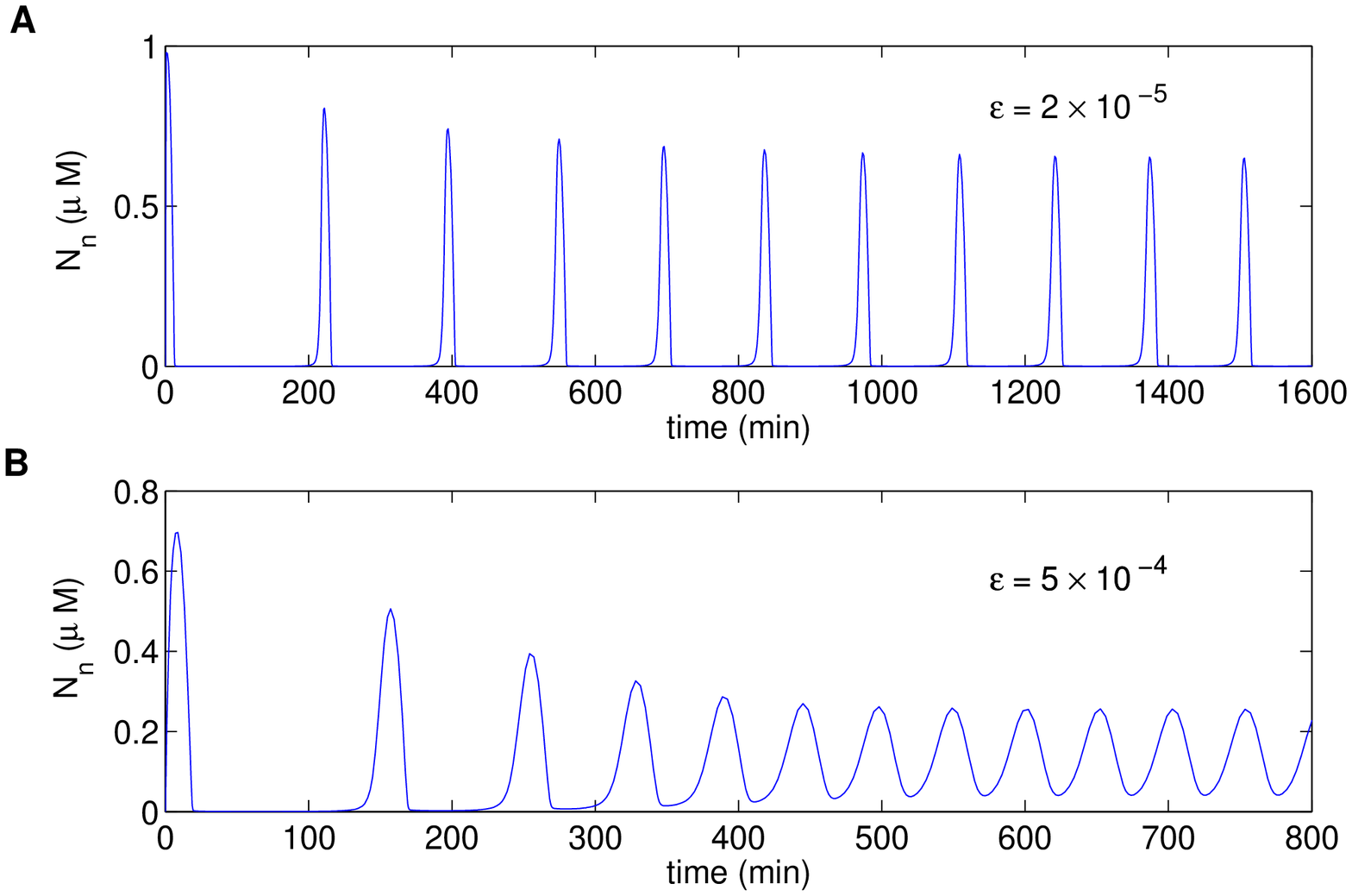}}
\centerline{\includegraphics[width=9cm]{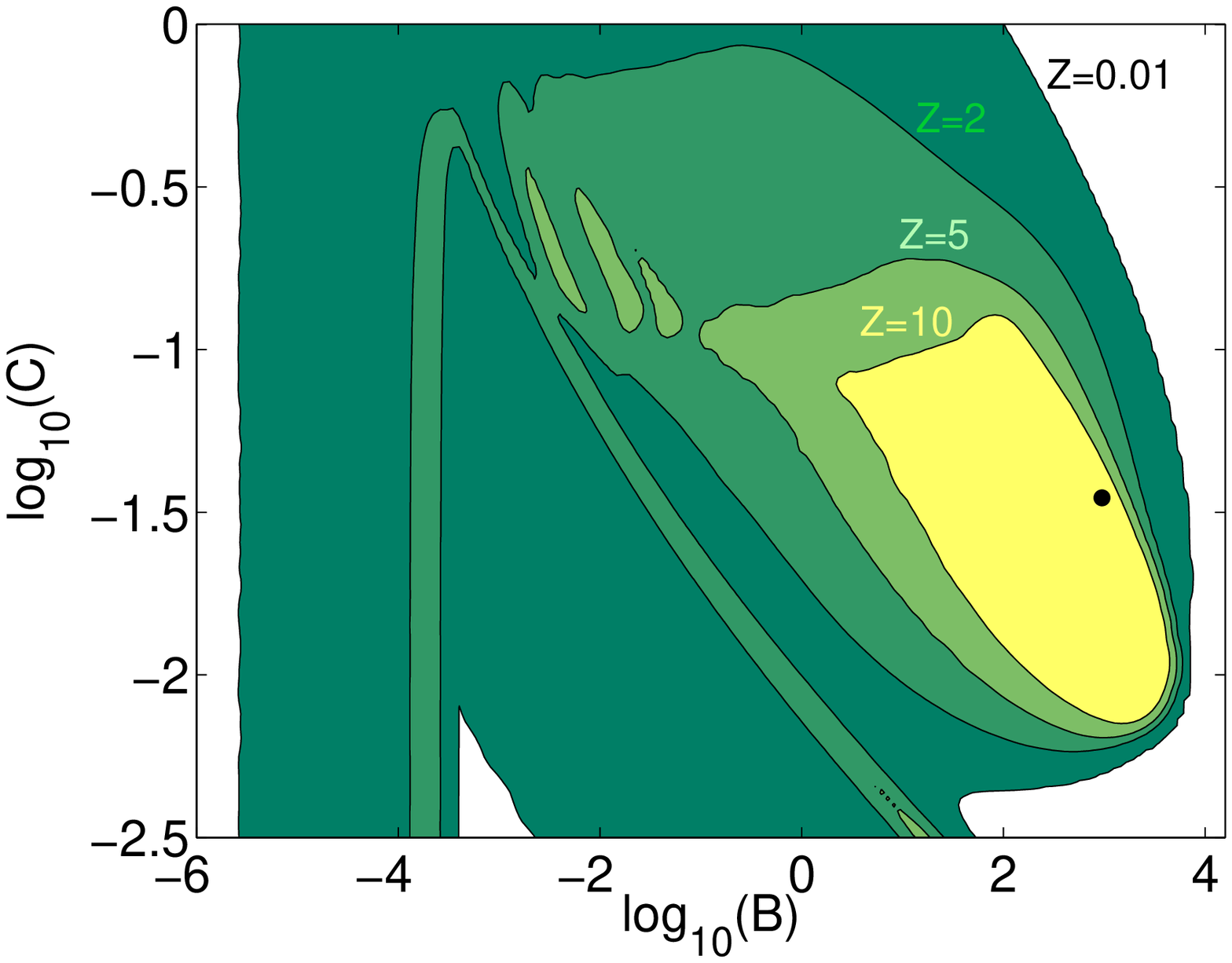}}
\caption{(A) Spiky oscillations in the NF-$\kappa$B model (parameter values are
identical to those used to produce Fig. \ref{osc}). (B) Soft oscillations, produced in the
same model using a larger value of $\epsilon$, keeping all other parameters unchanged. 
(C) A countour plot of the $Z$ measure of spikiness (see text), in the $B-C$ parameter plane
(other parameters are unchanged).
The $Z=0.01$ contour maps, approximately, the region of oscillations. The $Z=2$ contour
shows the region of spiky oscillations. The black dot marks the value of $B$ and $C$
used in (A) and (B), as well as in Fig. \ref{osc}.} \label{spikyandsoft} \end{figure}

One property of the oscillations of
nuclear NF-$\kappa$B, in figure \ref{osc}, that stands out is that
they are extremely spiky.
By choosing different parameter values one can also get soft oscillations
(see Fig. \ref{spikyandsoft}a,b), however, for biologically relevant
parameters the oscillations are spiky.
We suggest the following measure of spikiness:
\begin{equation}
Z=\frac{\mathrm{max}-\mathrm{min}}{\mathrm{average}}
\label{spikiness}
\end{equation}
In other words, $Z$ is the ratio of the amplitude of the oscillations
to the average level. Then, we will call a particular oscillation
{\em spiky} if $Z>2$ and {\em soft} otherwise. Further, $Z=0$ indicates
the absence of oscillations.
Fig. \ref{spikyandsoft}c shows a contour plot of $Z$ values on
the $B-C$ plane. In some directions the system transitions quickly from
no oscillations to spiky oscillations, wheras in other directions there
is a softer transition.
In general, the existence and spikiness of the oscillations is very robust
to changes in most of the parameters of the model \cite{sandeeppnas}.

\begin{figure}[t] {\bf A\hfill B\hfill ~}\\ \centerline{
\includegraphics[width=6.5cm]{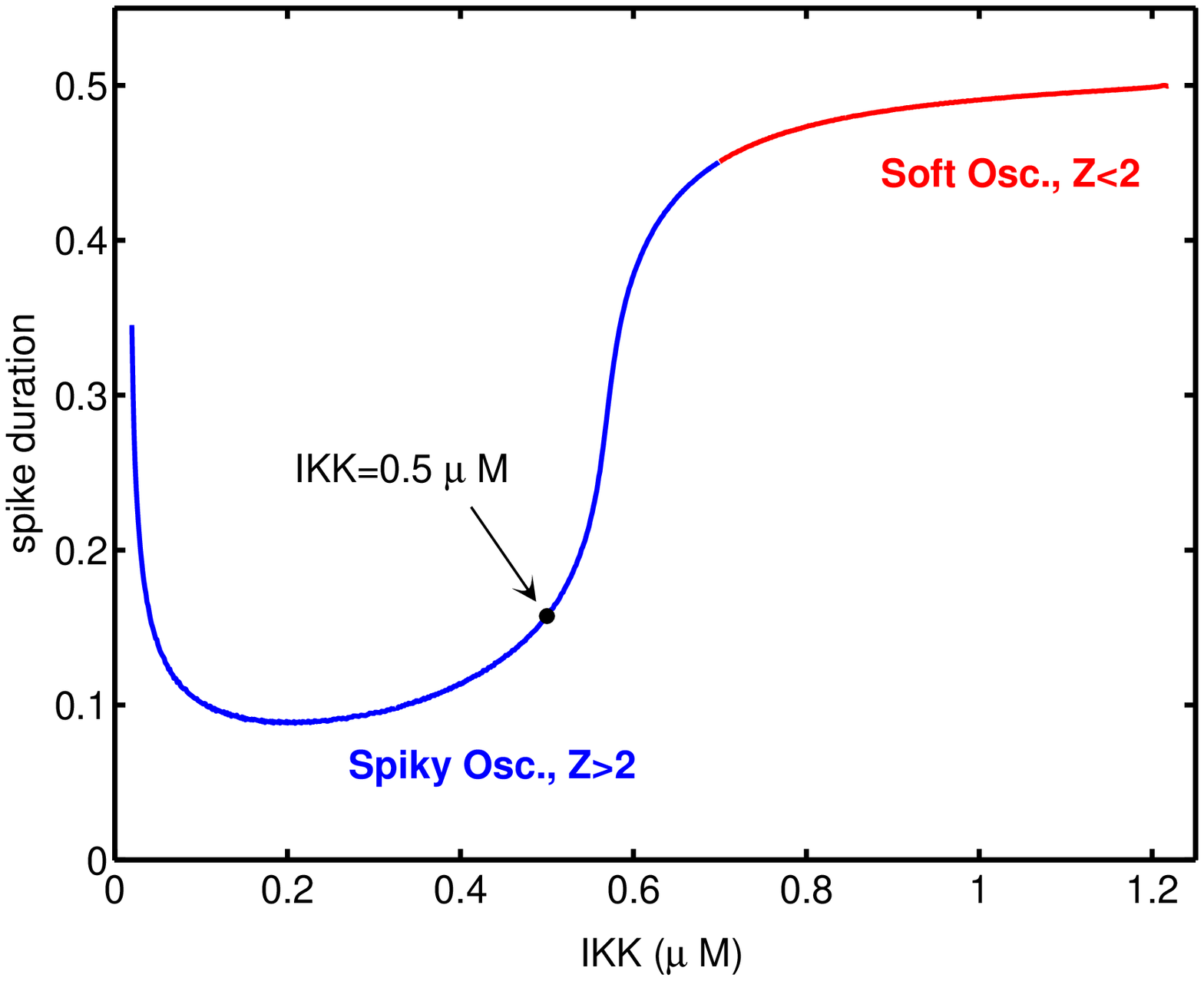}
\includegraphics[width=6.5cm]{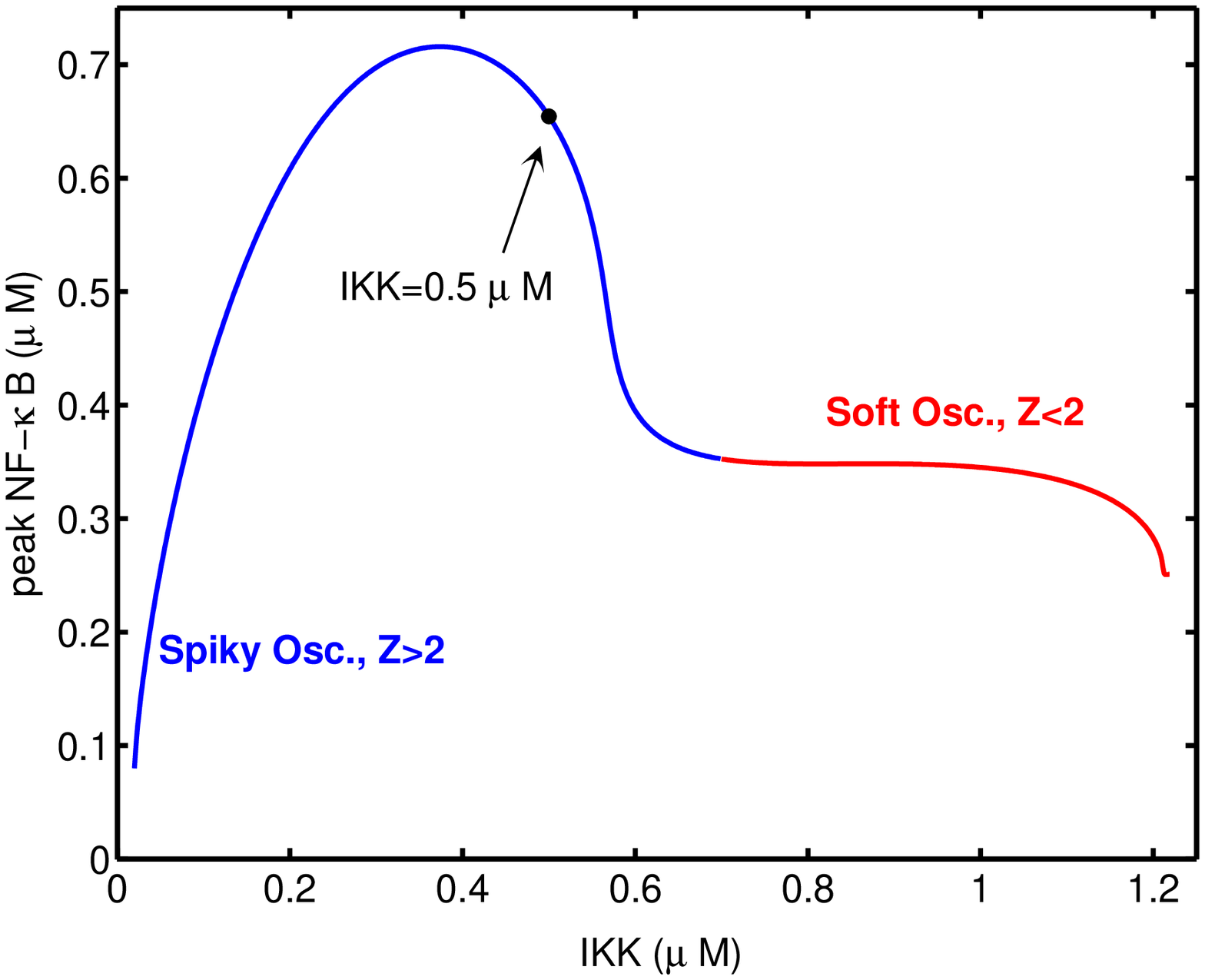} }
\caption{\label{IKKsensitivity} Sensitivity to IKK.  {\bf A.} Spike duration,
the fraction of time $N_n$ spends above its mean value, as a function of IKK
concentration. 
{\bf B.} Spike peak, the maximum concentration of nuclear NF-$\kappa$B, as a function of
IKK concentration. 
In both plots,
the black dot shows the IKK value used in Fig. \ref{osc}, while
blue and red signify, respectively, regions of spiky and soft oscillations.}
\end{figure}

There is one parameter, however, to changes in which the system shows a very 
sensitive response. That parameter is the external input, IKK.
Fig. \ref{IKKsensitivity} shows that both the spike height (or peak level),
as well as the spike duration, can change by large amounts in response to
small changes in IKK level. Notice that this sensitivity is particularly high in
IKK ranges which are near the transition from spiky to soft oscillations.
One way of quantifying this sensitivity is to measure the expression level
of a gene whose transcription is activated by NF-$\kappa$B.
Imagine a gene whose upstream regulatory region contains a binding
site for NF-$\kappa$B dimers. Upon, binding the gene promoter is activated: 
$ G + 2N \mathop{\rightleftharpoons}_{k_{off}}^{k_{on}} G^*.$ 
Experimental measurements of NF-$\kappa$B-dependent gene expression suggest
that many genes closely follow the oscillations of NF-$\kappa$B \cite{Bosisioetal}.
This corresponds to the case where the 
binding of NF-$\kappa$B to the operator is in equilibrium,
i.e., $k_{on}$ and $k_{off}$ are much larger than the rates of all other
processes in the NF-$\kappa$B system. In that case the gene activity, $G^*$,
will follow the NF-$\kappa$B concentration:
$G^*=\frac{N_n^2}{k_{off}/k_{on}+N_n^2}$.
In this case the downstream gene inherits the high sensitivity of
the NF-$\kappa$B signal to IKK: the peak gene activity as a function
of IKK is a very steep sigmoidal curve with an effective Hill coefficient
of over 20 \cite{sandeeppnas}. Such a large value is very unusual for
biological systems, and is much larger than the values
obtained by other mechanisms \cite{HF,GK}.  

\section{A tool to analyse oscillation patterns in time series}

After having discussed the mechanism underlying the production of 
oscillations and studied in detail three examples of ``ultradian" oscillations
in cells, we would like to describe an easy tool
\cite{maybepnas} capable of providing information on the architecture of the
underlying network, given the experimental time series data for
oscillating biological systems.

More precisely, given the time sequence of maxima and minima of the
concentrations of the species during the oscillations, the method allows one to
assess whether the observed sequence is compatible with a
a single underlying negative feedback loop, describable by Eq. \ref{generalN}. 
If it is, the method further 
predicts the logical structure of the loop, i.e.
the order of the proteins within the loop, as well as which protein acts as an
activator and which as a repressor.

The method is grounded in a mathematical result valid for all monotone systems
describable by Eq. \ref{generalN}. For such a system, we can prove
the following statements about the sequence of maxima and minima in the time series:
\begin{itemize} 
\item the extrema (maxima or minima) have to
follow the order of the variables; for example, after a maximum of variable
$i-1$ there can either be a maximum or a minumum of variable $i$, and 
\item if two successive extrema are of the same kind (both maxima, or both minima), then
variable $i$ is activated by variable $i-1$. Conversely, if they are of
opposite nature (one maximum and one minimum), then variable $i$ is repressed
by variable $i-1$.  
\end{itemize}

\begin{figure} 
\centerline{\includegraphics[width=10cm]{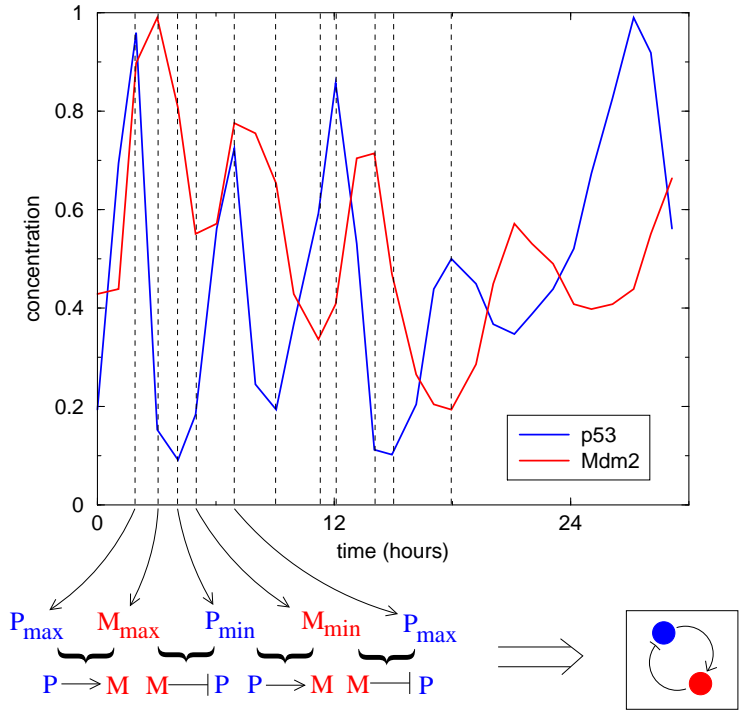}}
\caption{Reconstructing the underlying loop from the time series. The time series
shows p53-Mdm2 oscillations observed in single cell fluorescence experiments \protect\cite{GevaZatorsky}.
Using the rules mentioned in the text, the sequence of maxima and minima is converted
to the feedback loop shown at bottom right.} \label{algorithm} \end{figure}

\begin{figure} 
\centerline{\includegraphics[width=10cm]{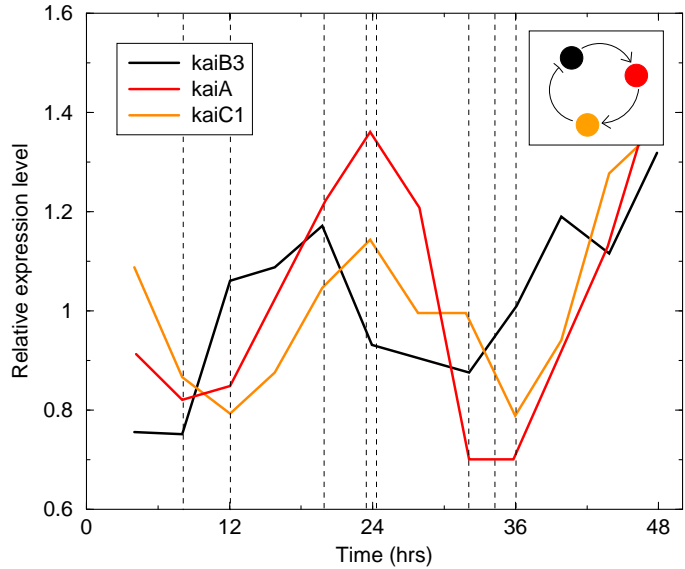}}
\caption{Reconstruction algorithm applied to circadian rhythms in cyanobacteria. Data from \protect\cite{kuchocircadian}.
The algorithm predicts that {\em kaiA} activates {\em kaiC1}, which represses {\em kaiB3}, which, in turn, 
activates {\em kaiA}.} \label{cyano} \end{figure}

These statements can be used to reconstruct the structure of the underlying
loop from the time series, if it is compatible with the above restrictions.
We illustrate this using the example of p53-Mdm2 time series shown in Fig. \ref{algorithm}.
The reconstructed loop is exactly the structure we used in Eq. \ref{eq_p53}: p53
activates Mdm2, which represses p53.
Of course, in this case, there are only two possible loop structures, and
the correct interactions are already known experimentally so the information gained is minimal.
The method comes into its own for systems with more variables. One example
is the cyanobacterial circadian clock. Fig. \ref{cyano} shows oscillations in the expression
of three circadian clock genes in {\em Synechocystis} and the reconstructed loop.
Two of the interactions have been experimentally observed in another strain
of cyanobacteria \cite{Golden_JohnsonKondo}, but the activation of {\em kaiA} by {\em kaiB3} is a new prediction.

Notice that there are many possible maxima/minima sequences which are not
compatible with the above rules. For example, in order to be compatible with a
single loop, each species must have exactly one maximum and one minimum during
one period. Moreover, if the method predicts an even number of repressors, one
should consider an oscillating positive feedback loop, which is impossible (see
\cite{maybepnas} for a full list of non--allowed cases). In all these
situations, one should conclude that the mechanism causing the oscillations is
more complicated and cannot be reduced to a single feedback loop.
Even if the structure of the real protein interaction network is more
complicated that a single loop (which is usually the case), the method can 
hint at the interactions which are most relevant for the oscillating
mechanism and help in building up a zero-order model of the system.

Finally, we note that while the method is mathematically rigorous for systems without time delays,
it works even if there are unobserved chemical species taking part to
the loop \cite{maybepnas}. As intermediate species introduce time delays it is likely that the result can be
extended to systems with an explicit time delay added to Eq. \ref{generalN}.

\section{Summary and Outlook}

Questions about cellular processes that use complex, temporally varying
signals can be crudely classified into the `How' and `Why' groups.
'How' questions deal with the structure of the regulatory network
and the range of dynamical behaviour it can produce: how does the network
produce oscillations? what are the necessary and sufficient mechanisms?
how are they implemented in real cellular systems?
`Why' questions deal with the physiological role of the particular
dynamical behaviour produced by the network: why does p53 start oscillating
in response to DNA damage? do oscillations carry some information that can
be decoded by downstream genes? could a non-oscillating system have worked
equally well for the cell? 

In this review we have mainly investigated the
`How' questions for one subset of cellular
response systems that use temporally varying signals:
eukaryotic systems exhibiting
``ultradian" oscillations.
The three systems we have modelled -- p53, which is important
for apoptosis, Hes1, which is part of the Notch cycle
responsible for somite segmentation, and NF-$\kappa$B, a key protein
in immune response -- all turn out to have the same basic design that produces
oscillations. The two key aspects of this design are the presence of
{\em negative feedback} and {\em time delays}.
There are many ways of producing an effective time delay in cellular
processes. In particular, we emphasised, using the example of NF-$\kappa$B,
{\em saturated degradation} 
as one such mechanism.

Saturated degradation plays a role in many models of oscillatory negative
feedback--loops. The earliest use of this mechanism is in the model
of Bliss, Painter and Marr \cite{BPM}, which is similar to
our NF-$\kappa$B model. 
Saturated degradation has also been used by Goldbeter 
in various models of cellular oscillations, e.g.,
the cell cycle  \cite{Goldbeter_cellcycle}, development in myxobacteria
\cite{Goldbeter_development}, yeast stress response
\cite{Goldbeter_yeaststress} and the mammalian circadian clock
\cite{Goldbeter_circadian}. It is also found in models of
calcium oscillations  in cells \cite{RBSSZE,Goldbeter_calcium}.
In the p53-Mdm2 system too, as described earlier, there is a
similar saturated complex formation, where one component
of the complex is subsequently degraded. This is so similar to
the NF-$\kappa$B case that it should be possible to construct
a model for p53 oscillations which has ordinary differential
equations without an explicit time delay (see for instance \cite{GevaZatorsky}).
Because saturated degradation is easily
implemented by complex formation we conclude that it would be a very
useful general mechanism for producing an effective time delay in
cellular systems.

Although we have focussed more on how the oscillations are produced,
the `Why' question of 
whether the oscillations have a direct physiological role in
these systems is of course an important one.
In NF-$\kappa$B, oscillations are observed, not in wild-type cells, but
rather in mutants or cells which overexpress certain proteins.
Therefore, it is quite possible that oscillations themselves
play no direct role in the wild-type response \cite{BWKCHL}.
However, wild-type cells do show a complex temporal variation of NF-$\kappa$B
of which the spiky response of the I$\kappa$B$\alpha$ module is an
extremely important component (I$\kappa$B$alpha$ is the only isoform
whose knockout mutants are not viable \cite{baltimore}). 
Hes1 oscillations, even though damped, might have a direct relevance
for spatial pattern formation in embryos. Hes1 is 
a key element in the Notch oscillating cycle, which, in mouse
and chicken embryos, is coupled, out of phase,
with the Wnt cycle. In each period of these cycles, 
a somite is formed, thus leading to vertebrate segmentation.
Further work should clarify whether the Hes1 oscillations drives
these cycles, or is slaved to them, or has a different role entirely.
In the case of p53, the oscillations appear to be sustained, therefore
it is likely they are directly important for the apoptosis pathway,
but again it is not clear precisely how.

Even in the absence of unambiguous evidence that oscillations have a direct
functional importance, 
investigations into the mechanisms underlying the oscillations 
have implications for some of the `Why' questions. 
For instance, noting that an oscillating signal carries much more information than a steady signal,
\cite{Lahav,TE} suggested that differential regulation of downstream gene circuits could be achieved if
they were sensitive to the frequency of the oscillation. Indeed, as we have shown \cite{sandeeppnas},
it is easy for a single downstream gene to act as a `low-pass filter'. Regulatory
circuits with multiple genes could be constructed to have particular frequency responses.
However, it is perhaps more fruitful to look for a physiological role
in other properties of the dynamics, rather than the periodicity. 
One such property, especially prominent in NF-$\kappa$B oscillations, is the
spikiness.
A spiky signal of this kind carries even more information than a soft oscillation
or a steady state level. 
Spiky pulses, whether
periodic or random or isolated, are in a sense less 
``expensive": If a downstream gene is expressed when p53 crosses some threshold it
is energetically and metabolically cheaper to have a spike whose 
peak crosses the threshold for long enough to trigger the gene,
than to produce and maintain the protein above that level
for a long time. 

In this context it is noteworthy that several of the most ubiquitous signalling
molecules in eukaryotes, hormones, exhibit recurrent spikes.
The spikes sometime come in
a nearly periodic fashion but may also appear more randomly. 
Why hormones appear in spikes is also not known. Again, we speculate that
spikes are an efficient way to trigger other regulatory mechanisms
in the body as compared to a steady level which do not deliver `kicks'
to stimulate other compartments. 
An added benefit is that the time between spikes allows an
extra level of regulation.
In addition, it might not be good
for a biological organism to be subjected to a constant level of
hormones all around the clock. 

A related issue where spikiness has implications is that of cross-talk between signals. 
Thus, a high constant level of a protein may potentially interfere with other signals
being sent, for instance by saturating common receptors.
In contrast, a `quantization' of the signal into spikes allows different
signals to use common receptors without fear of interference.
An analogy to cars is perhaps useful to visualize
this difference: If there were no traffic lights, it would still be
easy to drive a car to your destination provided the traffic was not
too heavy. However, if all cars were made 10 times longer it would
become much harder to drive at the same density of traffic.
Indeed, railway tracks have less intersections because trains are,
in effect, very long cars.

On a more concrete level, our analysis of the NF-$\kappa$B network suggests
that the spikiness of the oscillations was correlated to a sharp,
threshold-like response of the system.
The output (nuclear NF-$\kappa$B concentration), we found, could be extremely sensitive
to changes in the input (IKK) while remaining relatively insensitive
to other parameters; a clear prediction that awaits experimental verification.
Such a threshold-like response is likely to
be very important for the differential regulation of downstream genes.
It would be very interesting to know if there is a causal connection between
spikiness and sharp responses in this system, as well as other
regulatory systems.

Overall, we conclude that oscillations, especially spiky ones,
have many useful properties
that could be exploited to improve the speed and
efficiency of signalling and response systems. 

\vspace{0.5cm}

{\it We are grateful to Alexander Hoffmann, Terry Hwa,
Arnie Levine, Eric Siggia, Galit Lahav and Ian Dodd
for interesting discussions on oscillating genetics. S. K., M. H. J and K. S. acknowledge support from The Danish National Research Foundation and Villum Kann Rasmussen Foundation. G. T. acknowledge support from the FIRB 2003 program of the Italian Ministry for University and Scientific Research.}




\appendix

\section{Properties of monotone systems}

In this section we study the fixed point properties of
a feedback loop composed of an arbitrary number, $N$, of nodes
whose dynamics is given by Eq. (\ref{generalN}) in the main text, which we repeat here:
\begin{equation}
\label{eq:model_s}
\frac{dx_i}{dt}=g_i^{(A,R)}(x_i, x_{i-1})\qquad i=1\ldots N.
\end{equation}
                                                                                                                                
Our analysis proceeds by noting that, using the monotonicity
condition, we can write explicit functional relations between
neighboring variables in the steady state (when $dx_i/dt=0$):
                                                                                                                                                                                                                                                                 
\begin{equation}
g_i^{(A,R)}(x_i^*,x_{i-1}^*)=0\quad\Rightarrow\quad  x_i^*=f_i^{(A,R)}(x_{i-1}^*)
\end{equation}
Notice that the functions $f_i$ have the same monotonicity
properties as the $g_i$s with respect to the second
argument (for this it is necessary that $g_i(x,y)$ be a monotonically {\em decreasing} function of $x$).
By iterative substitution, we obtain:
\begin{eqnarray}\label{fixed2_s}
x_i^*=f_i(x^*_{i-1})=f_i(f_{i-1}(x^*_{i-2}))=\ldots=\nonumber\\
=f_i\circ f_{i-1}\circ f_{i-2}\circ\ldots\circ f_{i+1}(x^*_i)\equiv F_i(x^*_i)
\end{eqnarray}
where $\circ$ denotes convolution of functions. Here, we introduced
the function $F_i(x)$, which quantifies how the species $i$ interacts
with itself by transmitting signals along the loop.  Notice also that
if Eq.(\ref{fixed2_s}) holds for one value of $i$, then it holds for any
$i$, since it is sufficient to apply $f_{i+1}()$ on both sides to
obtain the equation for $x_{i+1}^*$ and so on. For feedback loops,
much useful information can be obtained from the properties of
$F_i(x)$.  Firstly, by applying the chain rule, we obtain the slope of
$F_i(x)$ at $x$:
$F_i'(x)=\prod_j f_j'(x_j)|_{x_i=x}$.  The r.h.s is always greater
(less) than zero if the number of repressors present in the loop is
even (odd). In the former case, there can be multiple fixed points, i.e.,
this is a necessary condition for multistability.
On the other hand, when there are an odd number of repressors, then $F_i(x)$ is positive and monotonically
decreasing, meaning that there is one and only one solution to the
fixed point equation $x^*_i=F_i(x_i^*)$.

To perform the stability analysis, we write the characteristic
polynomial evaluated at the fixed  point:
\begin{equation}
\prod_i \left[\lambda-\partial_x g_i(x,y)|_{x=x^*}\right]=\prod_i \partial_y g_i(x,y)|_{x=x^*}.
\end{equation}

The above equation can be greatly simplified using the relation
$F'(x)=\prod_i \partial_y g_i(x,y)/\partial_x g_i(x,y)$, which is a
consequence of the implicit function theorem and the chain rule.  One
then obtains the following equation:
\begin{equation}\label{linstab_s}
\prod_{i=1}^N\left(\frac{\lambda}{h_i}+1\right)=F'(x^*)
\end{equation}
where the $h_i=- \partial_x g_i(x_i,x_{i=1})|_{x^*}$ are the
degradation rates at the fixed point. Notice that, because $F'(x)$ is always
negative in a negative feedback loop, all coefficients of the characteristic polynomial
are non-negative, hence it can not have real positive
roots. This means that the destabilization of the fixed point can only
occur via a Hopf bifurcation, i.e. with two complex conjugate
eigenvalues crossing into the positive real half-plane.

\begin{figure}[ht]
\begin{center}
\includegraphics[width=7cm]{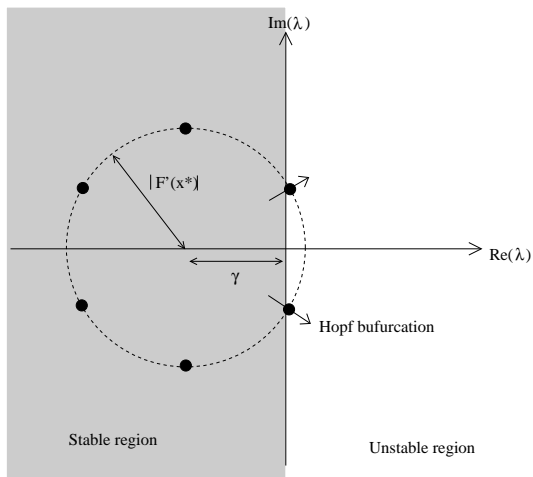}
\end{center}
\caption{Sketch of the Hopf bifurcation in the eigenvalue complex plane,
in the case in which all the degradation rates are equal to a constant $\gamma$.\label{figure3_s}}
\end{figure}

In the simple case in which all the degradation
rates are equal and unchanging (i.e, $h_i=\gamma$, a constant) the roots of the polynomial (\ref{linstab_s}) in
the complex plane are the vertices of a polygon centered on $-\gamma$ with a
radius $|F'|$ as sketched in Fig. \ref{figure3_s}.
Therefore, the fixed point will remain stable as long as
\begin{equation}
|F'(x^*)|\cos(\pi/N)<\gamma.
\label{stabcond}
\end{equation}
In this case, Hopf's theorem (see Appendix \ref{app_hopf}) ensures the existence of a periodic orbit
close to the transition value, whose period is:
\begin{equation}
  \label{eq:hopfperiod_s}
  T=2\pi/Im(\lambda)
\end{equation}
which, in the simple case of equal degradation timescales, becomes
$T=2\pi/[|F'(x^*)|\cdot \sin(\pi/N)]$.
Notice that the Hopf theorem does not ensure that the orbit is stable;
however, since the system is bounded and there are no other fixed
points, we expect the orbit to be attracting, at least close to the
transition point.

Notice that condition (\ref{stabcond}) is always satisfied when $N=2$.
This result also extends to the more general case where degradation rates are unequal.
Thus, we have proven that a two-component monotone negative feedback loop
without an explicit time delay can never show oscillations.

\section{Hopf bifurcations} \label{app_hopf}

Usually, the bifurcation takes place under the variation of some external
'control' parameter $\mu$ and the new state appears at a critical value $\mu_c$
of this parameter.  For genetically regulated networs, the control parameter
could be some external chemical stimuli, a production rate, or a binding
constant, just to mention a few examples.  To illustrate a Hopf bifurcation let
us first consider a completely general two-dimentional dynamical system
(without delay) defined by two coupled non-linear differential equations
\begin{equation} \left( \begin{array}{c} \dot p \\ \dot m\end{array} \right)
~=~ \left( \begin{array}{c} f(p,m) \\ g(p,m)\end{array}\right) \end{equation}
(we use variables $(p,m)$ to resemble p53 and mdm2).  The stationary fixed
point $(p^*,m^*)$ is determined by \begin{equation} f(p^*,m^*) = g(p^*,m^*) = 0
\end{equation} Using standard routines one linearizes around the fixed point by
means of the Jacobian matrix \begin{equation} J ~=~\left(\begin{array}{cc}
\frac{\partial f}{\partial p} & \frac{\partial f}{\partial m} \\ \frac{\partial
g}{\partial p} & \frac{\partial g}{\partial m} \end{array} \right)_*
\end{equation} where the star symbolizes that we insert the fixed point into
the Jacobian. The resulting set of eigenvalues $\lambda_1,\lambda_2$ can either
both be real or be a set of two complex conjugates. This is the case we are
interested in here: \begin{equation} \lambda_1= \alpha + i\omega~~,~~\lambda_2=
\alpha - i\omega \end{equation} For $\alpha < 0$ (correpsonding to $\mu <
\mu_c$) the fixed point $(p^*,m^*)$ is stable whereas the limit cycle becomes
stable for $\alpha > 0$ (corresponding to $\mu > \mu_c$) and performing
oscillations on the limit cycle with a frequency $\omega$ defining th period of
the oscillation.  For a genetic system this period could for instance be
circadian, ultradiand or related to cell cycles.  Just above the bifurcation,
the limit cycle can be well approximated by a circle whoes radius for
super-critical Hopf bifurcations generally grows continuosly as $\sqrt{\mu -
\mu_c}$ (as in contrast to a sub-critical Hopf bifurcation where the radius
will jump and exhibit hysteresis effects).
When the control parameter $\mu$ is much larger than $\mu_c$ the limit cycle
may easily be deformed in various ways 'away' from the circle. 

When we introduce a time delay in the equations, as discussed above, we can
formally write it by introducing the delayed variable $p_\tau$ into the
equations: \begin{equation} \left( \begin{array}{c} \dot p \\ \dot m\end{array}
\right) ~=~ \left( \begin{array}{c} f(p,m) \\ g(p,m,p_\tau)\end{array} \right)
\end{equation} We find numerically that the system still exhibits Hopf
bifurcations when crossing critical lines in the parameter space. In fact, we
find that one can drive the system through Hopf bifurcations by increasing the
value of the time delay $\tau$ through specific values.

The fixed point of $p$ is of course equal to the fixed point of $p_\tau$,
i..e.: $f(p^*,m^*) = g(p^*,m^*,p^*) = 0$. Now we use the same approach as above
and linearize around the fixed point: \begin{equation}
q=p-p^*,r=m-m^*,q_\tau=p_\tau-p^* \end{equation} which leads to the following
dynamical equations for the increments: \begin{eqnarray} \dot q &=&
f(p^*+q,m^*+r) = f(p^*,m^*) + \frac{\partial f}{\partial p} \cdot q +
\frac{\partial f}{\partial m} \cdot r=\nonumber \\ &=& 0 + \alpha \cdot q +
\beta \cdot r \end{eqnarray} \begin{eqnarray} \dot r &=&
g(p^*+q,m^*+r,p^*+q_\tau) = \nonumber \\ &=& g(p^*,m^*,p^*) + \frac{\partial
g}{\partial p} \cdot q + \frac{\partial g}{\partial m} \cdot r + \frac{\partial
g}{\partial p_\tau} \cdot q_\tau= \nonumber \\ &=& 0 + \gamma \cdot q + \delta
\cdot r + \epsilon \cdot q_\tau \end{eqnarray} We can now assume solutions on
the form: \begin{equation} q(t)=q_1 e^{\lambda_1 t} + q_2 e^{\lambda_2 t}, ~~~~
r(t)=r_1 e^{\lambda_1 t} + r_2 e^{\lambda_2 t} \end{equation} and find after
some lengthy algebra the eigenvalues: \begin{equation} \lambda_1 = \frac{1}{2}
\left[ \alpha + \delta \pm ((\alpha - \delta)^2 + 4 \beta (\epsilon
e^{-\lambda_1 \tau} + \gamma))^{\frac{1}{2}} \right] \end{equation}
\begin{equation} \lambda_2 = \frac{1}{2} \left[ \alpha + \delta \pm ((\alpha -
\delta)^2 + 4 \beta (\epsilon e^{-\lambda_2 \tau} + \gamma))^{\frac{1}{2}}
\right] \end{equation} We note that these are transcendal equations in the
eigenvalues $\lambda_1, \lambda_2$ but that the time delay $\tau$ specifically
appears into the relations (for a further discussion, see previous Appendix).

In many biological systems with genetic feed back regulations, there are often
more than two dynamical variables.  We shall in the following sections discuss
oscillatory states of the transcription factor NF-$\kappa$B. As we show, that system
can favourably be reduced to a three dimensional dynamical system in the
variables NF-$\kappa$B, its inhibitor I$\kappa$B and the associated mRNA, I$\kappa$Bm. Let us for
simplicity write these variables as $N,I,I_m$ thus formally obtaining the
following three dimensional dynamical system: \begin{equation} \left(
\begin{array}{c} \dot N \\ \dot I_m \\ \dot I\end{array} \right) ~=~ \left(
\begin{array}{c} f(N,I_m,I) \\ g(N,I_m,I) \\ h(N,I_m,I)\end{array}\right)
\end{equation} The procedure to study this type of quations is exactly the same
as for the two-dimensional system. From the stationary point $N^*,I_m^*,I^*$ we
linearize around it using the Jacobian matrix.  \begin{equation} J
~=~\left(\begin{array}{ccc} \frac{\partial f}{\partial N} & \frac{\partial
f}{\partial {I_m}} & \frac{\partial f}{\partial I} \\ \frac{\partial
g}{\partial N} & \frac{\partial g}{\partial {I_m}} & \frac{\partial g}{\partial
I} \\ \frac{\partial h}{\partial N} & \frac{\partial h}{\partial {I_m}} &
\frac{\partial h}{\partial I} \end{array} \right)_* \end{equation} In this case
one obtains three eigenvalues $\lambda_1,\lambda_2,\lambda_3$ which are either
all three real, or one is real and the two others complex conjugates. Again, in
the last case one may encounter a Hopf bifurcation where the fixed point
$N^*,I_m^*,I^*$ bifurcates into a limit cycle in the three-dimensional
$N,I_m,I$ space, causing oscillations in these variables. Notice, that there
topologically is a big difference between oscillatory limit cycles in two and
three dimensions. In two dimensions a limit cycle can 'enclose' trajectories,
which is not the case in three dimensions where tracjectories may 'escape
around' the limit cycles. An important theorem for this type of behavior is
called the Poincare-Bendixson theorem, which states that if a trajectory is
confined to a closed, bounded region and there are no fixed points in that
region, then the trajectory eventually approaches a closed orbit (see e.g.
\cite{strogatz}).

\section{Stability analysis of delay systems}

\subsection{Linear delay systems} \label{app_linear}

Consider the simplest delayed system of the form \begin{equation}
\frac{dx}{dt}=ax(t)+bx(t-\tau).  \label{eq_app_linear} \end{equation} A
complete description of the solution is given in an appendix of ref.
\cite{glass}, while we will just investigate the conditions where the system
display sustained oscillations. The general solution of Eq.
(\ref{eq_app_linear}) is $x=\exp(\lambda t)$ which, inserted in t investigate
the conditions where the system display sustained oscillations. The general
solution of Eq. (\ref{eq_app_linear}) gives \begin{equation}
\lambda=a+be^{-\lambda\tau}.  \end{equation} Defining $\lambda=\mu+i\omega$,
the oscillatory case takes place when $\mu=0$. The solution is then
\begin{equation} i\omega=a+b\cos\omega\tau-ib\sin\omega\tau \end{equation}
which gives \begin{eqnarray} \omega&=&(b^2-a^2)^{1/2}\\
\tau&=&\frac{arccos(-a/b)}{(b^2-a^2)^{1/2}}.  \end{eqnarray} A more general
treatment \cite{hayes50} gives the conditions under which the system converges
to a stationary solution, that is $|a|>|b|$ for any $\tau$ or if $|a|<|b|$ for
$\tau<arccos(-a/b)/(  (b^2-a^2)^{1/2} )$. Roughly speaking, the system can
oscillate if the dominant part of the kernel is the delayed one and if the
delay is large enough.

\subsection{Absence of closed orbits} \label{app_dulac}

If we label the vector forming the left--hand of Eq. (\ref{general}) as ${\bf
x}$ and that forming the side right--hand side as ${\bf F}$, its divergence of
${\bf F}$ is \begin{equation} \nabla\cdot{\bf F}=\frac{\partial f}{\partial
x}-k_x+\frac{\partial g}{\partial y}-k_y.  \end{equation} Since we exclude that
each of the two molecules can activate itself, the partial derivatives
$\partial f/\partial x$ and $\partial g/\partial y$ are non--positive, and
consequently $\nabla\cdot{\bf F}\leq0$. By virtue of Green's theorem, if the
trajectory $C$ were closed, \begin{equation} 0>\int \nabla\cdot {\bf\dot
x}dA=\oint_C {\bf\dot x}\cdot{\bf n}\;dl, \end{equation} where ${\bf n}$ is the
normal to the trajectory, and consequently ${\bf \dot x\cdot n}=0$ everywhere.
Since the circulation of a null vector is zero, this leads to a contradiction,
and the trajectory cannot be closed.

\subsection{Stability analysis of the p53-mdm2 system} \label{app_p53}

Neaumtu and coworkers develop in ref. \cite{neamtu} the complete stability
analysis of the p53-mdm2 system with delay. First they show that for any choice
of the parameters there is a unique stationary point for the rate equations.
Consider the system obtained by linearization of Eq. (\ref{eq_p53}) aroud such
point \begin{eqnarray} \frac{\partial p}{\partial t} & = &
(a\rho_{10}-b)p(t)-a\rho_{01}m(t)\nonumber\\ \frac{\partial m}{\partial t} & =
& -d m(t) +c\gamma_{10}p(t-\tau)+c\gamma_{01}p(t-\tau), \end{eqnarray} where
$\rho_{10}$ and $\rho_{01}$ are the derivatives of $pm(t)$ with respect to $p$
and $m$ in the stationary point, while $\gamma_{10}$ and $\gamma_{10}$ are the
derivatives of the function $p-pm/(k_g+p-pm)$. Define $p_1=b+d+a\rho_{10}$,
$p_0=db+ad\rho_{10}$, $q_1=c\gamma_{01}$ and
$q_0=c\gamma_{01}(b+a\rho_{10})-ac\rho_{01}\gamma_{10}$. Let $\lambda$ be the
eigenvalues of the linearized system.  It is possible to prove that if $\tau>0$
and $p_1^2q_0^2-q_1^2p_0^2-2p_0q_0^2>0$ then there is a delay $\tau_0$ such
that \begin{equation}
Re\left(\frac{d\lambda}{d\tau}\right)_{\lambda=i\omega_0,\;tau=\tau_0}>0
\end{equation} and consequently a Hopf bifurcation occurs at the stationary
point.

In the limit of large dissociation constant $k$, $\rho_{10}=\rho_{01}=0$.
Consequently the condition for a Hopf bifurcation begins $b^2>0$, which is
always satisfied (except for the trivial case $b=0$).

The critical delay $\tau_0$ is given by \begin{equation}
\tau_0=\frac{1}{\omega_0}\left(arcsin\frac{p_1\omega_0}{\sqrt{(p_0-\omega_0^2)^2+\omega_0^2p_1^2}}+
arcsin\frac{q_1\omega_0}{\sqrt{(p_0-\omega_0^2)^2+\omega_0^2p_1^2}} \right),
\end{equation} where $\omega_0$ is the solution of the equation
\begin{equation} \omega^4+(-p_1^2-2p_0+q_1^2)\omega^2+p_0^2-q_0^2=0.
\end{equation}
  

\end{document}